\documentclass[12pt]{article}
\usepackage[centertags]{amsmath}
\usepackage{fancyhdr}
\usepackage[bottom]{footmisc}

\usepackage{amsfonts}
\usepackage{amssymb}
\usepackage{amsthm}
\usepackage{newlfont}
\usepackage[english]{babel}
\usepackage{graphicx}
\usepackage{array}
\usepackage{enumitem}

\usepackage{titlesec}

\titleformat{\section}
{\normalfont\large\bfseries}
{\thesection. }
{0pt}
{}

\numberwithin{equation}{section}

\begin{document}

\def\baselinestretch{1.7}

\title{\textbf{The  $\boldsymbol{SO(2r)_2}$  string functions as $\boldsymbol{q}$-diagrams} \vspace{-0.1in}}
\author{Arel Genish and Doron Gepner \vspace{0.1in}\\
\small\textit{{Department of Particle Physics, Weizmann Institute, Rehovot, Israel}}   \vspace{-0.1in} \\
\small\textit{{E-mail: Arel.genish@weizmann.ac.il, Doron.gepner@weizmann.ac.il}}
\date{\small{\today}}}

\maketitle

{\center ABSTRACT \\}

 We discuss our conjecture for simply laced Lie algebras level two string functions of mark one fundamental weights and prove it for the $SO(2r)$ algebra. To prove our conjecture we introduce $q$-diagrams and examine the diagrammatic interpretations of known identities by Euler, Cauchy, Heine, Jacobi and Ramanujan. Interestingly, the diagrammatic approach implies these identities are related in the sense that they represent the first few terms in an infinite series of diagrammatic identities. Furthermore, these diagrammatic identities entail all the identities needed to prove our conjecture as well as generalise it to all $SO(2r)$ level two string functions. As such, our main objective is proving these series of diagrammatic identities thus extending the works mentioned and establishing our conjecture for the $SO(2r)$ level two string functions.

\section{\large{Introduction}}

Over time, physics and mathematics have developed a fruitful interchanging relationship. A prominent example is the fascinating connection between characters of CFT, RSOS models and number theory \cite{lep, das, ked, bavgep, kun, belgep}. This connection was first noted by Baxter \cite{bax} in his work regarding the Hard Hexagon model where he found that the local state probability is governed by a one dimensional configuration sum. Interestingly, in the appropriate regime, this one dimensional sum corresponded on one hand to the sum side of the famous Roger Ramanujan identity \cite{wad}, while on the other, to characters of a fixed point CFT. This relationship was considerably further developed in later works \cite{mel, gep} and leads to the conjecture that $q$ sums identities exist for every CFT that appears as a fixed point for some RSOS model in some regime, and vice versa. \\
Our research focusses on parafermionic conformal field theories associated with Lie algebras ref. \cite{gep1} which were analogously developed in mathematics as $Z$-algebras \cite{lep1}.
These theories are described by cosets of the type,
\begin{equation}\label{-2.1}
H(G_r)=\frac{G_r}{U(1)^r},
\end{equation}
where $G_r$ is any Lie algebra of rank $r$ and level $l$. Furthermore, denote the simple roots of $G_r$ by $\alpha_i$ and the fundamental weights by $\omega_i$ where $i=1,...,r$. The fields in the theory are labeled by a pair of weights $(\Lambda,\lambda)$, where $\Lambda$ is a dominant highest weight (DHW) of $G_r$ at level $l$ and $\lambda$ is an element of the weight lattice
of $G_r$. Additionally, for a non zero field, these weights follow the selection rule,
\begin{eqnarray}\label{-2.2}
\Lambda-\lambda=\beta,
\end{eqnarray}
where $\beta=\sum \beta_i\alpha_i$ belongs to the root lattice of $G_r$ denoted $M_G$, i.e integer $\beta_i$.
The dimension of these fields, up to an integer, is given by,
\begin{eqnarray}\label{-2.3}
h^{\Lambda}_{\beta}=\frac{(\Lambda,\Lambda+2\rho)}{2(l+\mathfrak{g})}-\frac{(\Lambda-\beta)^2}{2l}  \mod1,
\end{eqnarray}
where  $\mathfrak{g}$ denotes the dual Coxeter number while $\rho=\sum\omega_i$ is the Weyl vector. Note, that somewhat unconventionally we denote the dimension by $\Lambda$ and $\beta$ as this will prove more convenient. The characters of the coset theory encompass the descendent structure for the different primary fields. These characters, denoted by $H({G_r})^{\Lambda}_{\beta}$, produce the $G_r$ string functions\footnote{When multiplied by an appropriate eta function factor} which are of central importance in the study of Lie algebras and are the main interest of our recent papers. More specifically, for level two simply laced Lie algebras, exact expressions for all characters were found in ref. \cite{gengep} via the ladder coset construction. In addition, a conjectured expression for $H(G_r)^{\Lambda}_{\beta}$, for some cases of $\Lambda$, was given and verified numerically. Furthermore, in recent work by Gepner \cite{gep2}, a generalisation for all Lie algebras at any level, was given and verified numerically .  \\
This paper is devoted to proving our conjecture for the characters of $H(SO(2r))$ at level two. Let us recall the conjecture for $H({G_r})^{\Lambda}_{\beta}$ where $G_r$ is any simply laced Lie algebras at level two \cite{gengep}. Define the $G_r$ root,
\begin{eqnarray}\label{-2.3.1}
Q=\beta \mod 2M_G,
\end{eqnarray}
where $M_G$ denotes the root lattice of $G_r$ and $Q=\sum Q_i\alpha_i$ such that $Q_i=0,1$. Additionally, we introduce the $q$ Pochhammer symbol
\begin{eqnarray}\label{-2.3.2}
(a,q)_n=\begin{cases} \prod\limits^{n-1}_{l=0}(1-aq^l) & n > 0 \\ 
1 & n=0 \\
\prod\limits^{-n-1}_{l=0}  1/(1-aq^{-1-l}) & n < 0 \end{cases}, 
\end{eqnarray}
which we will often abbreviate $(a,q)_n\equiv (a)_n$. \\
Our conjecture for the characters associated with $\Lambda$ zero or a fundamental weight of mark\footnote{The mark of $\omega_i$, denoted $a_i$, is defined by $a_i=\theta\omega_i$ where $\theta$ is the highest root of $G_r$} $1$ is given by,
\begin{eqnarray}\label{-2.4}
H(G_r)^{\Lambda}_{\beta}=q^{-d_{Q}^{\Lambda}}\sum_{ \{b_i\}=0\atop  {b_i= Q_i \mod2} }^
\infty 
{q^{ b^2/4-b\Lambda/2 }\over (q)_{b_1}...(q)_{b_r}},
\end{eqnarray}
where $b=\sum b_i\alpha_i$ is a root, we sum over $b_1,b_2,...b_r$ even or odd  according to the restriction $b_i=Q_i \mod2$ and $d^{\Lambda}_{Q}$ is some dimension which we do not specify for now.  Indeed, for $\Lambda=0$, this has been a long standing conjecture \cite{kun} dating back to $1993$ where it was originally motivated by TBA considerations. As discussed in \cite{gengep} our conjecture encapsulates all previous conjectures as well as providing new ones.  \\
Our main objective here is to prove our conjecture for the $SO(2r)$ algebra at level $2$.  As shown, in ref. \cite{gengep}, the $H(SO(2r))$ coset theory at level two is equivalent to a $\mathbb{Z}_2$ orbifolded theory of $r-1$  bosons moving on a $\sqrt2 M_{SU(r)}$ lattice. The characters of such a theory, are well known and are given in section ($4$). As such, combined with our expression eq. (\ref{-2.4}), they provide an infinite number of conjectured $q$ sums identities. The appearance of such identities highlights the underlying connection between characters of CFT and number theory and proving them is the main objective of this paper. \\
To prove these identities we first present $q$-diagrams. Viewed from number theory perspective, their study will highlight a connection between some well known identities as well as provide an interesting interpretation of the simply laced Lie algebras level $2$ string functions as diagrammatic extensions of these famous works. 
These extended identities enable us to prove the mentioned identities as well as find similar identities for any DHW character. Thus, extending our expressions to all characters of the $H(SO(2r))$ coset theory and providing a vast number of new identities.

\section{\large{$\boldsymbol{q}$-diagrams}}

In the introduction we have described our conjectured identities for the $SO(2r)$ string functions. How could one go about proving such identities? The purpose of this section is to introduce $q$-diagrams as a general tool for the proof of such identities. As usual with diagrammatic notations, at first sight they might seem as nothing but an elegant way of writing long sums. Indeed, often diagrammatic notations are only as good as the intuition they provide. Such intuition arises from the following two observations. First, $q$-diagrams can be shown to possess symmetries which could be realised as Weyl symmetries. This statement will be made explicit in the next section.  Second, the diagrammatic rules which arise trivially from our conjecture can be used to describe the classical identities of Euler, Cauchy, Ramanujan, Jacobi, Heine and more  \cite{RRS}. Beautifully, the diagrammatic expressions imply these identities, discovered by different mathematician years apart, are related. More specifically, they represent only the first terms of some infinite series of such identities. Following this diagrammatic intuition, to be discussed in section ($4$), the rest of our paper is devoted to proving these infinite series of identities thus extending the works mentioned. Finally, we will find that the conjectured identities for the $SO(2r)$ string functions of fundamental weights of mark one are the simplest case of these new identities. Moreover, all the $SO(2r)$ string functions are given by our new identities thus, generalizing our results.   \\
Let us start by introducing $q$-diagrams, recall our conjecture for the level two string functions of any simply laced Lie algebras denoted $G_r$ eq. (\ref{-2.4}). To identify the diagrammatic rules needed to describe $H^{\Lambda}_Q$ write the $q$ power explicitly,
\begin{eqnarray}\label{-1.2}
\frac{1}{4} \sum^{r}_{i,j=1} b_iG_{ij}b_j-\frac{1}{2} \sum^{r}_{i=1}b_i\Lambda_i 
 \end{eqnarray}
where $G_{ij}$ is the Cartan matrix of $G_r$ represented via its Dynkin diagram. Using the $G_r$ Dynkin diagram we introduce a set of diagrammatic rules. First, 
assign for each node at the Dynkin diagram some "momenta" $b_i$ such that $i$ corresponds to the number of the node. In addition, assign a momenta $\Lambda_i$ for each external line connected to the $i$'th node. Next, we prescribe a set of diagrammatic rules, 
\begin{enumerate}[label=\roman{*}., ref=(\roman{*})]
\item \text{for each node} $=\ {q^{b_{i}^2/2}\over (q)_{b_i}}$
\item \text{for each internal line connecting the $i$'th and $j$'th nodes} $=q^{-b_ib_j/2}$
\item \text{for each external line of momenta $\Lambda_{i}$ connected to the $i$'th node} $=q^{-\Lambda_ib_i/2}$
\item \text{sum over all nodes momentas }
$=\sum_{ b_i= 0 }^\infty\frac{1}{2}(1+(-1)^{b_i+Q_i})$
\end{enumerate}
where, for now, let us consider $\Lambda=\sum \Lambda_i\omega_i$ any weight with integer Dynkin labels  greater or equal to zero while $Q=\sum Q_i \alpha_i$ is any root vector of $G_r$.
Using this notation, our conjecture for the $G_r$ level $2$ string functions is simply given by the corresponding Dynkin diagram and two vectors Q and $\Lambda$. For 
example, consider the $q$-diagrams corresponding to the $SO(2r)$ Dynkin diagram, 
\begin{equation}\label{-1.4}
 \includegraphics[width=0.5\linewidth,keepaspectratio=true]{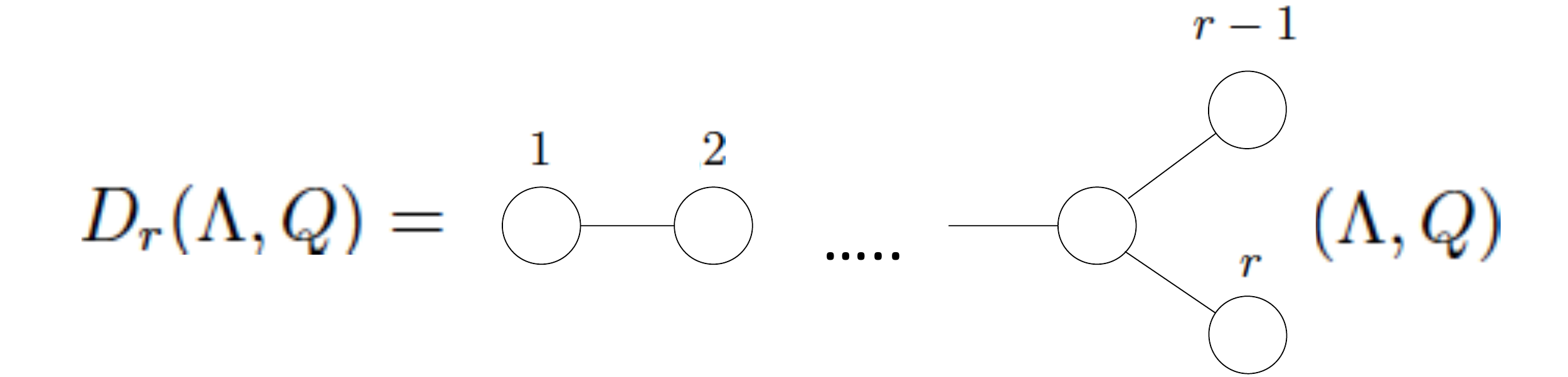}
  \end{equation}
where we denote the $q$-diagrams corresponding to $G_r$ by $G_r(\Lambda,Q)$ and label the nodes.  Then, the diagram contains $r$ nodes, $\Lambda$ is a $G_r$ weight specifying an assortment of external legs which we do not draw for now and $Q$ is a root of $G_r$ specifying parity restrictions on the summation. Following our diagrammatic rules these diagrams are given by
\begin{eqnarray}\label{-1.3}
D_r(\Lambda,Q)=\sum_{ b_i=0\atop  b_i= Q_i \mod2 }^
\infty 
{q^{ \frac{1}{2}(b_1^2-b_1b_2+b_2^2...+b_{r-2}^2-b_{r-2}(b_{r-1}+b_{r})+b_{r-1}^2+b_{r}^2-b_1\Lambda_1...-b_r\Lambda_r) }\over (q)_{b_1}(q)_{b_2}...(q)_{b_r}},
\end{eqnarray}
so the conjecture for the characters is simply $H(SO(2r))^{\Lambda}_Q = q^{d_{Q}^{\Lambda}}D_r(\Lambda,Q)$.
Let us highlight a few features that will prove helpful when calculating $q$-diagrams. First, obviously one is free to start by summing over any of the nodes momenta as this corresponds to summing over $b_i$ for some chosen $i$. However, once a certain node has been chosen all the lines connected to it must be taken into account as they carry a $b_i$ dependance. Thus, summing over a certain node corresponds to replacing the sub diagram containing this node and all lines connected to it by it's solution. The sums remaining are then given by the resulting diagram along with the solution for the chosen node. Indeed, it should be clear that the root $Q$  and weight $\Lambda$ associated with some diagram are always vectors of length equal to the number of nodes in the diagram. To clarify this remark, let us formally introduce the one node diagram connected to an arbitrary number of lines $l$,
\begin{eqnarray}\label{-1.5}
\includegraphics[width=0.37\linewidth,keepaspectratio=true]{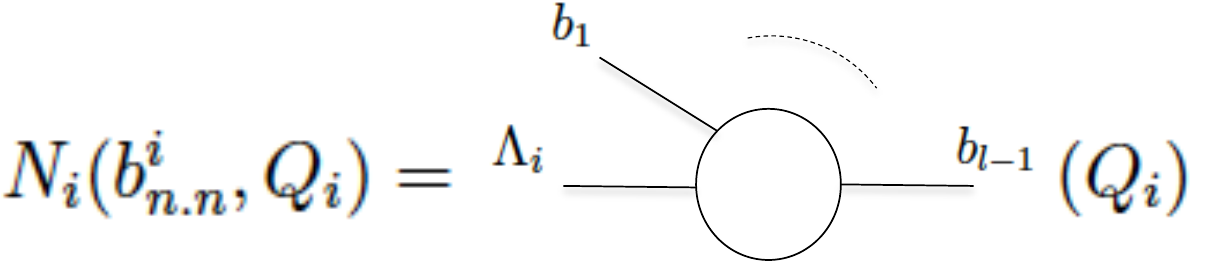}
\end{eqnarray}
where $b^i_{n.n}=\sum^{l-1}_{j=1}b_j+\Lambda_i$ denotes the connected lines momenta. Next, as a simple example, consider the $D_3(\Lambda,Q)$ diagram:
\begin{eqnarray}\label{-1.6}
\text{\includegraphics[width=0.75\linewidth,keepaspectratio=true]{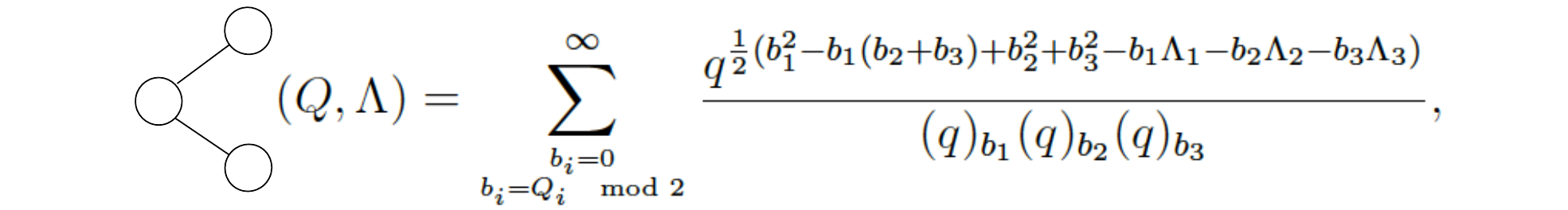}}
\end{eqnarray}  
where $Q=(Q_1,Q_2,Q_3)$ and $\Lambda=(\Lambda_1,\Lambda_2,\Lambda_3)$.
If we formally sum over the sub diagram containing only the third node with external lines carrying momenta $b_1$ and $\Lambda_3$  we find,
\begin{eqnarray}\label{-1.7}
\text{\includegraphics[width=0.9\linewidth,keepaspectratio=true]{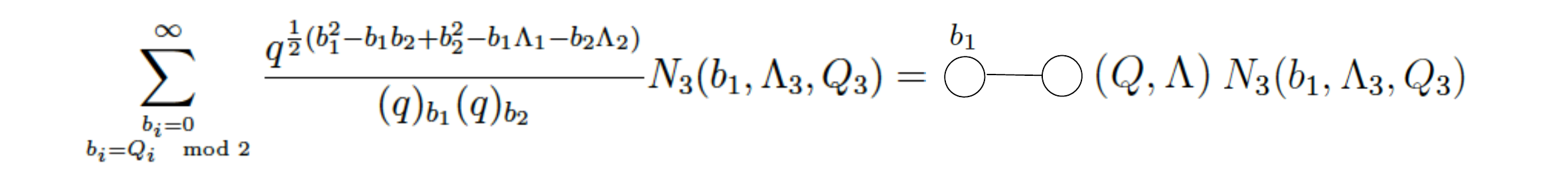}}
\end{eqnarray}  
Where now $Q=(Q_1,Q_2)$ and $\Lambda=(\Lambda_1,\Lambda_2)$ are vectors of length $2$ corresponding to the two nodes in the remaining diagram. 
Furthermore, note that once we have used the one node solution to solve the sum over $b_3$ we have a $b_1$ dependance in $N_3$. Thus, we can only use the one node solution to sum over $b_2$. Stated more generally, summing over the $i$'th node we get  $N_i(b_{n.n},\Lambda_i,Q_i)$ which carries a new dependance on the momenta corresponding to lines connected to the $i$'th node. Thus, we can only use the one node solution for non connected nodes. \\ 
As we proceed to calculate such diagrams many a times we shall consider sums of the same diagram albeit with different $Q$ roots. Naturally, this will be denoted in the following manner
\begin{eqnarray}\label{-1.8}
D(Q_1,\Lambda)+D(Q_2,\Lambda)+D(Q_3,\Lambda)...=D(\Lambda)(Q_1+Q_2+Q_3...),
\end{eqnarray}  
where $D$ stands for any diagram while $Q_i$ and $\Lambda$ are the roots and weight corresponding to $D$. Additionally, note that in such cases we can write the $Q$ dependance explicitly and  sum over the $Q$ dependent factors independently from the rest of the diagram,
\begin{eqnarray}\label{-1.9}
D(\Lambda)(Q_++Q_-)=D(\Lambda)\sum_{Q=Q_{\pm}}\frac{1}{2^r}(1+(-1)^{Q_1+b_1})(1+(-1)^{Q_2+b_2})...(1+(-1)^{Q_r+b_r})
\end{eqnarray}  
with $r$ being the number of nodes in $D$ and $Q_{\pm}$ are roots of length $r$.  \\
To further our study of $q$-diagrams we first present the one node diagram solution and study its symmetries.

\section{\large{$\boldsymbol{Q}$ symmetry and the Weyl group}}

In the previous section we have mentioned our conjecture is Weyl symmetric, let us take a closer look at this statement. Under a Weyl transformation $w$ our conjecture transforms as follows,
\begin{equation}\label{1.1}
H^{\Lambda}_{Q(\Lambda,w\lambda)}=H^{\Lambda}_{Q(\Lambda,\lambda)+\Delta_wQ}.
\end{equation}
Where $\Delta_wQ(\lambda)=(1-w)\lambda$. Although, $Q$ is defined modulo $2M_G$, where $M_G$ is the root lattice of $G$, $Q$ is not symmetric under the Weyl group. Consider the generators of the Weyl group denoted $s_i$,
\begin{equation}\label{1.2}
\Delta_{s_i}Q=\lambda_i\alpha_i,
\end{equation}
where $\lambda_i$ are the Dynkin labels of $\lambda$, i.e $\lambda=\sum \lambda_i \omega_i$. Clearly, for even $\lambda_i$ this is a symmetry of $Q$, however for odd $\lambda_i$ this is not a symmetry of $Q$. Thus, we should look for a $Q$ symmetry of $H^{\Lambda}_{Q(\Lambda,\lambda)}$ which compensates for this lack of symmetry in the definition of $Q$. To prove the symmetry eq. (\ref{1.1}) we study the $Q$ dependance of $H^{\Lambda}_{Q(\Lambda,\lambda)}$ through the one node diagram. This is actually a slightly more general setting, as for our conjecture $\Lambda$ is a fundamental weight of mark one and $Q=\Lambda-\lambda$ $\mod2M_G$ while these relations are not necessary for our definition of $q$-diagrams. \\
We now consider the single node diagram connected to an arbitrary number of lines $l$,
\begin{equation}\label{1.2.1}
\includegraphics[width=0.37\linewidth,keepaspectratio=true]{2}
\end{equation}
where $b^i_{n.n}=\sum_{j=1}^{l-1}b_j+\Lambda_i$ such that $b_j$ are some positive integers specifying an assortment of internal lines, $\Lambda_i$ is associated with the momenta of the external line and $Q_i$ is any integer.
Clearly any rank $r$ diagram, with its associated weight $\Lambda$ and root $Q$, can be constructed using $r$ $N_i$s. Recalling our diagrammatic rules, we can solve this one node diagram in the following way,
\begin{equation}\label{1.3}
N_i(b^i_{n.n},Q_i)=
\sum_{b_i=0\atop b_i=Q_i\mod2}^\infty\frac{q^{\frac{1}{2}b_i^2-\frac{1}{2}b_ib^i_{n.n}}}{(q)_{b_i}}=
\frac{1}{2}\sum_{b_i=0}^\infty\frac{q^{\frac{1}{2}b_i^2-\frac{1}{2}b_ib^i_{n.n}}}{(q)_{b_i}}(1+(-1)^{Q_i+b_i}).
\end{equation}
The sums appearing here are actually a combination of the Euler identity ref. \cite{RRS},
\begin{eqnarray}\label{1.4}
(-z)_{\infty}=\sum_{n=0}^\infty\frac{z^n q^{n(n-1)/2}}{(q)_{n}}.
\end{eqnarray}
From which we find,
\begin{eqnarray}\label{1.5}
N_i(b^i_{n.n},Q_i)=
\frac{1}{2}\biggl((-q^{(1-b^i_{n.n})/2})_{\infty}+(-1)^{Q_i}(q^{(1-b^i_{n.n})/2})_{\infty}\biggr),
\end{eqnarray}
clearly the $Q_i$ dependance is encompassed in the second term in the parenthesis. Since $(q^{-n})_{\infty}=0$ for any integer $n$  greater or equal to zero, we find the following $Q_i$ independence condition for positive $b^{i}_{n.n}$,
\begin{eqnarray}\label{1.6}
Q_{n.n} + \Lambda_i  \in 1+2\mathbb{Z}
\end{eqnarray}
where, recall that one of the lines is associated with an external line of momenta $\Lambda_i$ while the rest of the momenta are associated with internal lines. Accordingly,  we define $b_j=Q_j \mod 2$ and $Q_{n.n}=\sum_{j=1}^{l-1}Q_j$. As any $r$ node diagram can be constructed from the one node diagrams, we have thus proven 
\begin{eqnarray}\label{1.6.1}
Q_i \rightarrow Q_i+Q_{n.n}+\Lambda_i,
\end{eqnarray}
is a symmetry of any diagram including non negative external momenta. Let us study this symmetry for $q$-diagrams corresponding to Lie algebras.
Clearly, for  Lie algebra diagrams with external lines specified by $\Lambda_i$ and internal lines by $\sum_{j \neq i}G_{ij}b_j$, the $Q$ independence condition can be written as,
\begin{eqnarray}\label{1.7}
\Lambda_{i}-\sum_{j}G_{i,j}Q_j  \in 1+2\mathbb{Z}
\end{eqnarray}
where note that $G_{i,i}=2$. Let us examine this condition for the following definition of $Q$\footnote{One may note that other definitions of $Q$ can be considered. Indeed, to generalize our conjecture in section ($8$) we will consider a different definition.},
\begin{eqnarray}\label{1.8}
Q=\Lambda-\lambda \mod 2M_G
\end{eqnarray}
corresponding to our conjecture. Multiplying this definition by the simple root $\alpha_i$, 
\begin{eqnarray}\label{1.9}
\sum_{j}G_{i,j}Q_j=\Lambda_i-\lambda_i \mod 2.
\end{eqnarray}
Thus, the $Q_i$ independence condition can be written as,  
\begin{eqnarray}\label{1.11}
\lambda_i \in 1+2\mathbb{Z}.
\end{eqnarray}
This means that for some odd $\lambda_i$ our conjecture is independent of $Q_i$ i.e,
\begin{eqnarray}\label{1.12}
H^{\Lambda}_{Q(\Lambda,\lambda)}=H^{\Lambda}_{Q(\Lambda,\lambda)+\lambda_i\alpha_i}.
\end{eqnarray}
Clearly, this is also true for even $\lambda_i$ since $Q$ is defined only modulo $2M_G$. Moreover, this is exactly the transformation under the generators of the Weyl group $\Delta_{s_i}Q$ we found in eq. (\ref{1.2}). Thus, we have proven our conjecture is symmetric under the Weyl group. Furthermore, although our current interest lies in simply laced Lie algebras, it should be noted that Weyl symmetric $q$-diagrams can be constructed for any Lie algebra. \\
We can use the $Q$ symmetry of our conjecture (eq. \ref{1.12}) to find equivalent characters/diagrams with different $Q$ vectors, denoted equivalent $Q$ representations. These are build in a similar way to Lie algebras representations albeit mod $2$. Consider some character $H^\Lambda_Q$, to study its $Q$ dependance we look at $\lambda=\Lambda-Q$ Dynkin labels $\lambda_i$ mod $2$. For odd $\lambda_i$, we can use $s_i$ to shift $Q$ and $\lambda$ by $\alpha_i$ and find an equivalent character $H^\Lambda_{Q-\alpha_i}$. This procedure can be repeated for $\lambda-\alpha_i$ mod $2$ to find another equivalent character. Although this construction is quite parallel to that of Lie algebras representations of highest weight $\lambda$ there are a few subtleties. First, note that we are looking at $\lambda_i$ mod $2$, thus every weight is a highest weight. Second, when shifting $\lambda \rightarrow \lambda - \alpha_i$ mod $2$, we find that  $\lambda_i$ mod $2$ is not shifted. At first sight this implies that our representations are infinite, however shifting $\lambda$ we find an equivalent character with $Q_i \rightarrow Q_i +1$ mod $2$. Clearly, if we shift $\lambda$ by $\alpha_i$ again we just return to our original $Q$. Thus, to construct an equivalent $Q$ representation, we can consider only sequences of weights which do not include subtracting the same root twice in succession. Finally, this defines for us a lowest weight and insures that our representation is finite, i.e for a rank $r$ character there are $2^r$ different values for $Q$ and thus the representation dimension is bounded from above by $2^r$. Some examples in the following sections will clarify this procedure as well as facilitate the proof of our conjecture.

\section
{\large{Bosonic characters as $\boldsymbol{q}$-diagrams}}

The $SO(2r)_2$ string functions were found in our previous work \cite{gengep} via their correspondence to the $H(SO(2r)_2)$ coset characters. 
Using the ladder coset construction, we showed the coset theory is equivalent to a theory of $r-1$ free bosons $\mathbb{Z}_2$ orbifolded
and moving on a $\sqrt2 M_{SU(r)}$ lattice.
The fields, of such a theory, are labeled by $\lambda$, where $\lambda$ is a weight of $SU(r)$ and their dimensions are
\begin{equation}\label{2.-1}
\Delta_\lambda={\lambda^2\over 4}.
\end{equation}
In addition, there are the twist fields whose dimensions are $\Delta={r-1\over16},\ {r+7\over 16}$.
The characters are expressed by a level two $SU(r)$ classical theta
function defined by
\begin{equation}\label{2.0}
\Theta^r_{\lambda,m}(q)=\sum_{\mu\in M_{SU(r)}+\lambda/m} q^{\ m \mu^2/2},
\end{equation}
where $\lambda$ is any element of the weight lattice of $SU(r)$, $m$ is the level and $M_{SU(r)}$ denotes the root lattice of $SU(r)$.
Throughout this work we take the level to be $m=2$ as such we will omit the level index, i.e $\Theta^r_{\lambda,2}(q)=\Theta^r_{\lambda}(q)$. 
The characters of the $Z_2$ orbifold are divided into the twisted
and the untwisted sectors. These were given explicitly in ref. \cite{gengep}, here we will adapt the Pochhammer notation which will prove most useful. In the untwisted sector we have three types of characters. The characters of the
zero momenta fields are
\begin{equation}\label{2.1}
q^{(r-1)/24} \chi^r_\pm(q)={1\over2}\Theta^r_{0}(q) (q)_{\infty}^{-r+1}\pm
{1\over2}\ (-q)_{\infty}^{-r+1},
\end{equation}
where the second term arises due to the sector twisted in the time direction.
The nonzero momenta do not have such a contribution and are given simply by
\begin{equation}\label{2.2}
q^{(r-1)/24}\chi^r_\lambda(q)={1\over n_\lambda}{\Theta^r_{\lambda}(q)  (q)_{\infty}^{-r+1}},
\end{equation}
where $\lambda$ is a weight of $SU(r)$, $n_\lambda=2$ when $\lambda$ is on the root lattice and
$n_\lambda=1$, otherwise.
In the twisted sector, we have two fields which have the characters
\begin{equation}\label{2.3}
q^{(1-r)/48}\chi^r_{t,\pm}(q)={1\over2} (q^{1/2})_{\infty}^{-r+1}\pm
{1\over2}  (-q^{1/2})_{\infty}^{-r+1},
\end{equation}
where the upper index denotes the rank of the algebra.
We can determine which bosonic character corresponds to each coset character $H^{\Lambda}_{Q}$ by simply comparing (modulo one) the dimensions\footnote{Some care is required when there is more than one field of the same dimension} eqs. (\ref{-2.3}) and (\ref{2.-1}). As mentioned, not all of these characters are given by our conjecture, eq. (\ref{-2.4}), which specifies only characters corresponding to $\Lambda$ zero or a fundamental weight of mark one, i.e $\Lambda=\omega_i$ where $i=0,1,r-1,r$ and $\omega_0=0$. Calculating the corresponding dimension $h^{\omega_i}_Q$ it is easily verified that $H^{\omega_i}_Q$, where $i=r-1,r$ correspond to characters of the twisted sector while for $i=0,1,...,r-2$ they correspond to characters of the untwisted sector. Anticipating the results let us set,
\begin{equation}\label{2.3.1}
d^{\Lambda}_{Q}=\begin{cases} (r-1)/24 & \Lambda=0,\omega_1,...,\omega_{r-2} \\ 
(r-1)/48 &  \Lambda=\omega_{r-1},\omega_{r} \end{cases}
\end{equation}
additionally, in a similar manner to section ($2$), redefine $q^{(r-1)/48}\chi^r \rightarrow \chi^r$ for the twisted sector and similarly for the untwisted sector $q^{(r-1)/24}\chi^r\rightarrow \chi^r$. These redefinitions are a simple manner of convenience as we will find that these "renormalised" characters are simply given by the $SO(2r)$ $q$-diagrams. \\
With these results our conjecture for the characters implies some highly non-trivial $q$ sums identities. In what follows we consider only $\Lambda=0$ diagrams as such, to ease the discussion we omit $\Lambda$ from our notations. In section ($2$), we have mentioned that $q$-diagrams interpretations of some well known identities imply these identities are related which motivates some extensions of these works.  Additionally, these diagrammatic interpretations offer some intuition regrading our conjectured identities. Actually, such a $q$-diagrammatic interpretation was already encountered when solving the one node diagram. Indeed, we found that the Euler identity, eq. (\ref{1.4}), can be written as,
\begin{equation}\label{2.3.2}
\includegraphics[width=0.7\linewidth,keepaspectratio=true]{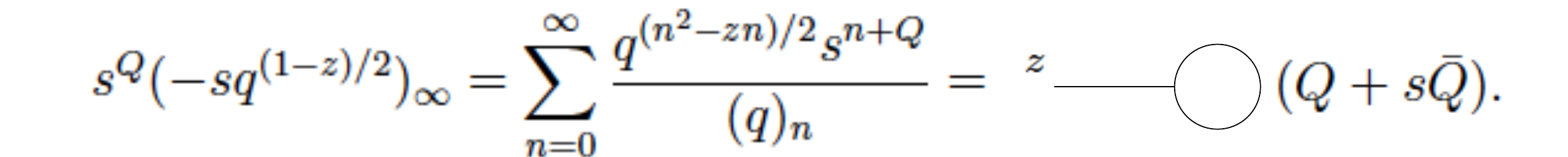}
 \end{equation}
Where with no loss of generality we replace $z$ with $-sq^{(1-z)/2}$ such that $s=\pm1$ and define $\bar{Q}=Q+1$. Additionally, we have used the notation introduced in the last section for a sum of the same diagram with different $Q$'s . Using this result, at $z=0$, it is easy to solve the $SO(4)$ diagram as it includes two non connected nodes,
\begin{equation}\label{2.3.3}
\includegraphics[width=0.28\linewidth,keepaspectratio=true]{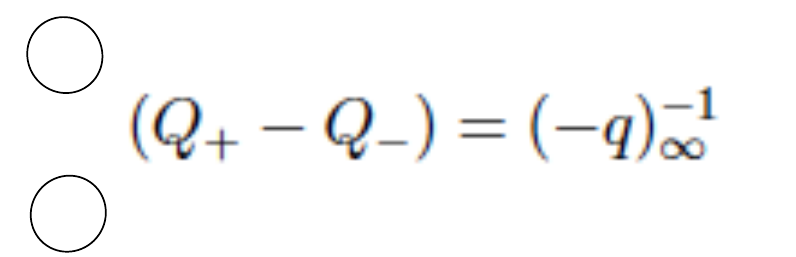}
\end{equation}  
where here $Q_+=(0,0)$, $Q_-=(1,1)$ and we have used $(-q)^{-1}_{\infty}=(-q^{1/2})_{\infty}(q^{1/2})_{\infty}$. \\
Next, consider the Cauchy identity,
 \cite{RRS},
 \begin{eqnarray}\label{2.3.4}
\frac{1}{(z)_\infty}=\sum_{n=0}^{\infty}\frac{z^nq^{n(n-1)}}{(q)_n(z)_n}.
\end{eqnarray}
At first sight, this identity does not seem to have a diagrammatic interpretation, however recall our discussion regrading solutions of $q$-diagrams. As we have highlighted, once some nodes are solved the diagrammatic interpretation might be "dressed" with some one node solutions. Indeed, as a simple example consider the case of $z=-q$. By using the Pochhammer identities,
 \begin{eqnarray}\label{2.3.5}
(q)_{2n}=(q)_{n}(-q)_{n}(q^{1/2})_{n}(-q^{1/2})_{n},   \hspace{0.4in}
(a)_{-n}(q/a)_{n}=(-q/a)^nq^{n(n-1)/2},
\end{eqnarray}
the Cauchy identity can be rewritten,
 \begin{eqnarray}\label{2.3.6}
\frac{1}{(-q)_\infty}=\sum_{n=0}^{\infty}\frac{q^{2n^2}}{(q)_{2n}(-q^{1/2})_{-n}(q^{1/2})_{-n}}.
\end{eqnarray}
To proceed let us introduce another basic Pochhammer identity,
 \begin{eqnarray}\label{2.3.7}
(aq^n)_{\infty}=\frac{  (a)_{\infty}     }{(a)_n}.
\end{eqnarray}
Dividing eq. (\ref{2.3.6}) by $(-q)_{\infty}$ and using this identity,
\begin{eqnarray}\label{2.3.7.1}
\frac{1}{(-q)^2_\infty}=\sum_{n=0}^{\infty}\frac{q^{2n^2}(-q^{1/2-n})_{\infty}(q^{1/2-n})_{\infty}}{(q)_{2n}}.
\end{eqnarray}
Finally, using the Euler identity of eq. (\ref{2.3.2}) to rewrite $(\pm q^{1/2-n})_{\infty}$ we find the Cauchy identity has the following $q$-diagram interpretation,
\begin{eqnarray}\label{2.3.8}
\text{\includegraphics[width=0.68\linewidth,keepaspectratio=true]{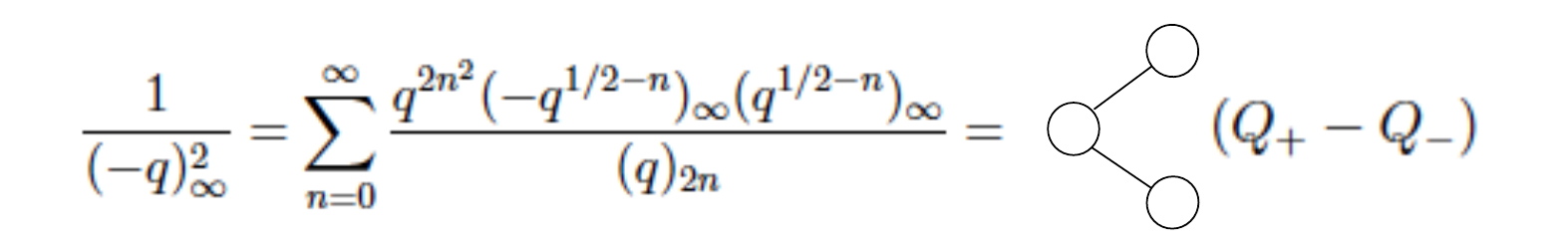}}
\end{eqnarray}
with $Q_+=(0,0,0)$ while $Q_-=(0,1,1)$. Remarkably, using $q$-diagrams it is evident that the Euler and Cauchy identities are related in sense that, their respective $q$-diagrams appear to be the first two in an infinite series of identities, 
\begin{equation}\label{2.3.9}
 \includegraphics[width=0.45\linewidth,keepaspectratio=true]{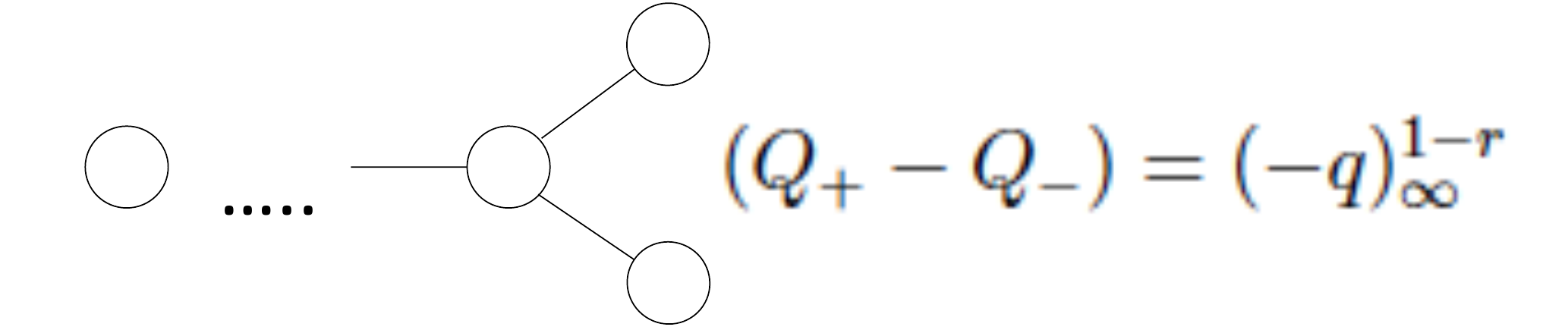},
 \end{equation}
where the diagram has $r$ nodes while $Q_+=(0,...,0)$ and $=Q_-(0,...,1,1)$ are rank $r$ roots\footnote{One may wonder about the $SO(8)$ diagram indeed via similar manipulations one finds this diagram is given by Heine's sum.}. It is easily verified, via dimension calculation \cite{gengep}, that the combination of diagrams appearing here are exactly our conjectured $SO(2r)$ diagrams for $\chi^{r}_+-\chi^{r}_-$. Additionally, following eq. (\ref{2.1}), it is clear that $(-q)^{r-1}_{\infty}$ corresponds to the contribution of the twisted sector in the time direction, thus implying our conjecture.  \\
At this point the reader might ask himself whether one can provide some diagrammatic intuition for the untwisted sector i.e, $(q)^{1-r}_{\infty}\Theta^r_0$. Such intuition is given by the following identities, due to Jacobi and Ramanujan \cite{RRS}, 
\begin{eqnarray}\label{2.3.10}
(q)^{-1}_{\infty}\sum_{n=-\infty }^{\infty}q^{n^2/2-nz/2}s^n=(-sq^{(1-z)/2})_{\infty}(-sq^{(1+z)/2})_{\infty},  \hspace{2in} \\
\label{2.3.10.1}
(q)^{-1}_{\infty}\sum_{n=-\infty }^{\infty}q^{3n^2/2-nz/2}s^n=\sum_{n=0}^\infty \frac{q^{n^2}}{(q)_{2n}}
  (-sq^{(1-z)/2})_{n}(-sq^{(1+z)/2})_{n},     \hspace{1.4in}         \\
(q)^{-1}_{\infty}(-q^{1/2})_{\infty}\sum_{n=-\infty }^{\infty}q^{n^2-nz/2}s^n=\sum_{n=0}^\infty \frac{q^{n^2/2}}{(q)_{n}(-q^{1/2})_n(-q)_{n}}
  (-sq^{(1-z)/2})_{n}(-sq^{(1+z)/2})_{n}.    
   \end{eqnarray}
The diagrammatic interpretation for these identities at $z=0$ can be achieved via similar manipulations as in our previous example. Elegantly, we again find that these diagrams are related. In the sense, that respectively they are given by the first three terms in following diagrammatic identity, 
\begin{eqnarray}\label{2.3.11}
\includegraphics[width=0.51\linewidth,keepaspectratio=true]{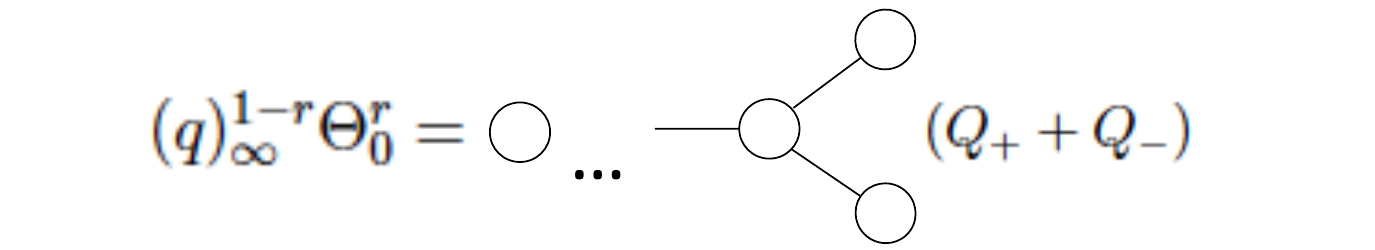}
\end{eqnarray}
where again the diagram contains $r$ nodes, $Q_{\pm}$ are defined as in eq. (\ref{2.3.9}) and the first diagram in the series is the non connected two node diagram corresponding to the Jacobi identity. Just as for the time twisted contribution, this extension of Jacobi and Ramanujan identities corresponds to our conjectured identities for the characters combination $\chi^{r}_++\chi^{r}_-$. \\
Clearly, for now, we have offered nothing but a diagrammatic motivation for our conjectured identities. However, the diagrammatic identities for the first terms in these series clarify their mathematical interpretation as diagrammatic extensions of works by Euler, Cauchy, Heine, Jacobi and Ramanujan. Furthermore, as mentioned in the introduction, these already represent new results. As, with the exception of $SO(4)$, the identities corresponding to $\Lambda=0$ proven here for the first few terms of the series  have been a long standing conjecture \cite{kun}. \\
Finally, our current discussion involved only $\Lambda=0$ diagrams, while our conjecture is also valid for $\Lambda=\omega_i$, where $i=1,r-1,r$. Using the diagrammatic notation these are given by the $SO(2r)$ Dynkin diagram with $\Lambda$ and $Q$. To ease the discussion we can encode $\Lambda=\omega_i$ in to our diagrammatic notation by filling the $i$'th node. Using this additional rule the different characters are given by the $SO(2r)$ Dynkin diagram with either no black nodes or one external black node and some $Q$.
Indeed, analogous diagrammatic arguments can be made for all our conjectured identities. The moral of this story is twofold. First, it is an interesting property that known identities seem to be related and this relation is made evident in the language of $q$-diagrams. Second, in our current discussion, diagrammatic interpretations were given to the Cauchy, Jacobi and Ramanujan identities at some specific value of $z$. A natural question is whether one can find diagrammatic interpretations for these identities at a general $z$. Following our diagrammatic rules, such diagrams must include external lines carrying the $z$ dependance. Indeed, diagrams with external lines are well suited for writing diagrammatic recursion relations. This motivates one to look for some diagrammatic recursion relations which will prove the extended series.  Accordingly, the next few sections which are devoted to proving these identities are constructed as follows.  First, we write recursion relations relating characters of rank $r$ to characters of lower rank, specifically $r-1$. Next, to prove our conjecture satisfies these recursion relations we study $D_3(\Lambda,Q)$ the three node diagram with external lines. Finally, we supplement these recursive proofs with the proof of an appropriate initial condition.

\section{\large{The twisted sector diagrams}}  

The conjectured identities which arise for the twisted sector were discussed in \cite{gengep}.  By calculating the dimension it can be shown that generally\footnote{For $r=4$ the diagram is symmetric to rotations of the external nodes} the twisted sector characters arise from $\Lambda=\omega_r,\omega_{r-1}$. These lead to the following conjectured identities,
\begin{equation}\label{2.3.1}
\includegraphics[width=0.37\linewidth,keepaspectratio=true]{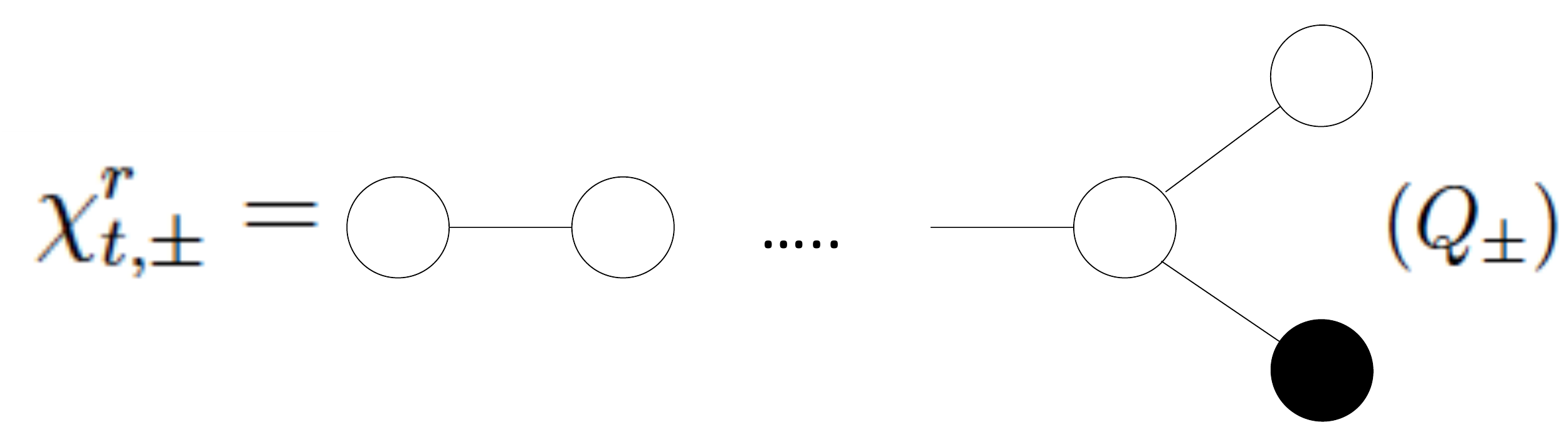}
\end{equation}
where $Q_{+}=(0,...,0)$ and $Q_{-}=(0,...,1,1)$. Clearly, our conjecture is symmetric under $\Lambda_r \leftrightarrow \Lambda_{r-1}$ and 
$Q_r \leftrightarrow Q_{r-1}$ so that similar identities hold for $\Lambda=\omega_{r-1}$ with $Q_{+}$ and $Q_{-}$. \\
To prove these identities we first write a recursion relation for the twisted sector characters of eq. (\ref{2.3}), 
\begin{equation}\label{2.4}
\chi^{r}_{t,s}={1\over2}\sum_{s'=\pm1}(s'q^{1/2})^{-1}_{\infty}(\chi^{r-1}_{t,s}+s'\chi^{r-1}_{t,\bar{s}} )
\end{equation}
where $s=\pm1$, $\bar{s}=-s$ and the upper index denotes the rank of the characters algebra. \\
To prove our diagrammatic expressions satisfy these recursion relations we look at the following $D_3$ diagram with an external line,
\begin{equation}\label{2.4.1}
\includegraphics[width=0.38\linewidth,keepaspectratio=true]{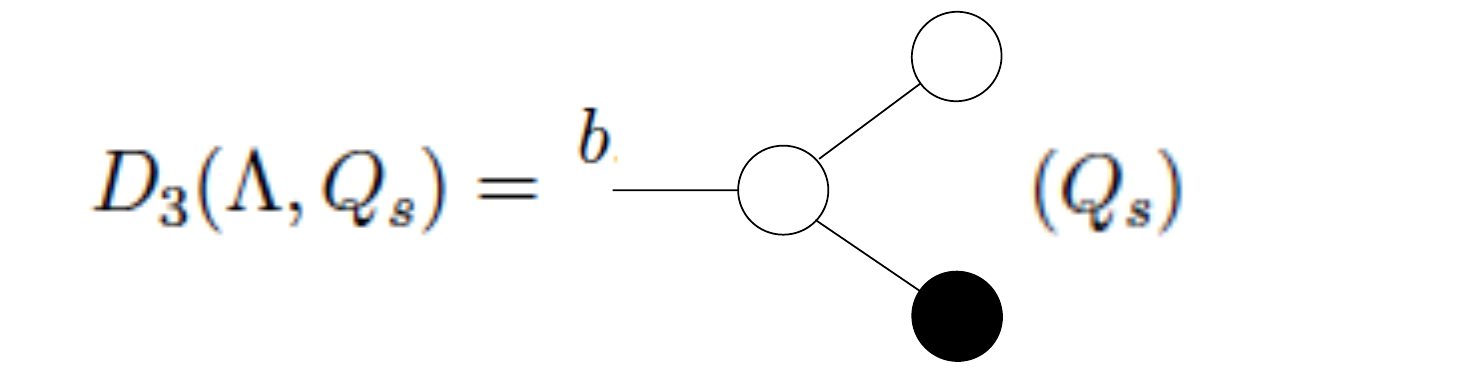}
\end{equation}
where  $\Lambda=b\omega_1+\omega_3$ and $Q_{s}=Q_{\pm}$. Clearly, for any rank $r$ the $D_r$ diagram includes this diagram, moreover we can use our expression for the one node diagrams to study the $D_3$ diagram. First, note that the one node diagram with $l$ external lines, eq. (\ref{1.5}), can be written as
\begin{equation}\label{2.5}
N_i(b_{n.n},\Lambda_i,Q_i)=
\frac{1}{2}\sum_{s_i=\pm1}  s_i^{Q_i}       (-s_iq^{(1-\Lambda_{i}-b_{n.n})/2})_{\infty}.
\end{equation}
If we now define $b_j=Q_j+2a_j$, for $j=1,..,l-1$, the $a_k$ dependance of this one node diagram can be extracted using the Pochhammer identity (eq. \ref{2.3.7}),
\begin{equation}\label{2.6}
N_i(b_{n.n},\Lambda_i,Q_i)=
\frac{1}{2}\sum_{s_i=\pm1}  s_i^{Q_i}      \frac{ (-s_iq^{(1-\Lambda_{i}-Q_k-\sum_{j \neq k}b_j)/2})_{\infty} }   { (-s_iq^{(1-\Lambda_{i}-Q_k-\sum_{j \neq k}b_j)/2})_{-a_k} }.
\end{equation}
As discussed in section ($2$), we can only use this solution of the one node diagram for non connected nodes. For our current diagram we use this solution for the two external nodes,
\begin{equation}\label{2.7}
N_{2}(b_1,0,Q_2)N_{3}(b_1,1,Q_3)=
\frac{1}{4}\sum_{s_{2}}         \frac{s_{2}^{Q_{2}} (-s_{2}q^{1/2})_{\infty} }   { (-s_{2}q^{1/2})_{-a_{1}} }
    \frac{ (-1)_{\infty} }   { (-1)_{-a_{1}} }
\end{equation}
here the contribution from $s_3=-1$ vanishes since $1-Q_{1}-\Lambda_3=0$ and $(1)_\infty=0$. With this expression $D_3$ is given by, 
\begin{equation}\label{2.8}
D_3(\Lambda,Q_{s})=
A \sum_{a_{1}=0}^{\infty}   \frac {q^{2a_{1}^2-a_1b}}  {(q)_{2a_{1}}   
       (-s_{2}q^{1/2})_{-a_{1}} 
        (-1)_{-a_{1}} }.
\end{equation}
Where $A$ denotes the $a_{2}$ independent terms,
\begin{equation}\label{2.9}
A=\frac{1}{4}\sum_{s_{2}} s_{2}^{Q_{2}} (-s_{2}q^{1/2})_{\infty}(-1)_{\infty}.
\end{equation}
Let us first simplify the denominator using the Pochhammer symbol identities,
\begin{eqnarray}\label{2.10}
 (q)_{2n}=(q)_{n}(-q)_{n}(q^{1/2})_{n}(-q^{1/2})_{n},   \hspace{0.4in}
(a)_{n}(q/a)_{-n}=(-a)^nq^{n(n-1)/2}.
\end{eqnarray}
So that the diagram is given by,
\begin{equation}\label{2.11}
D_3(\Lambda,Q_{s})=
A \sum_{a_{1}=0}^{\infty}   \frac {q^{a_{1}(a_{1}-b-\frac{1}{2})} s_{2}^{a_{1}} }  { 
   (q)_{a_{1}}     (s_{2}q^{1/2})_{a_{1}} }.
\end{equation}
To solve this sum replace $s_2q^{1/2}=y$ and note
\begin{equation}\label{2.11.1}
 (q)_{a_{1}}     (s_{2}q^{1/2})_{a_{1}}=(y^2,y^2)_{a_{1}} (y,y^2)_{a_{1}}=(y,y)_{2a_{1}}.
\end{equation}
The resulting sum is again solved using the Euler identity of eq. (\ref{1.4}),
\begin{equation}\label{2.12}
D_3(\Lambda,Q_{s})=
A \sum_{a_{1}=0}^{\infty}   \frac {y^{2a_{1}(a_{1}-b-\frac{1}{2})} s_2^{2a_1b}}   { 
        (y)_{2a_{1}} }=A\sum_{s_1=\pm1}(-s_1s_2^by^{-b})_{\infty}
\end{equation}
as this is just the one node sum in $y$. Additionally, for the time being we allow for non integer $b$. Finally, going back to $q$ 
\begin{equation}\label{2.12.1}
(-s_1q^{-b/2},s_2q^{1/2})_{\infty} =(-s_1q^{-b/2})_{\infty}(-s_1s_2q^{(1-b)/2)})_{\infty}
\end{equation}
and using $(-1)_{\infty}(-q^{1/2})_{\infty}(q^{1/2})_{\infty}=2$ we find the diagram is given by,
\begin{equation}\label{2.13}
D_3
(\Lambda,Q_{s})=
  \frac{1}{2}  \sum_{s_{1},s_2}  \frac{s_{2}^{Q_{2}} } {(s_{2}q^{1/2})_{\infty}}    (-s_{1}s_{2}q^{(1-b)/2})_{\infty} (-s_{1}q^{-b/2})_{\infty} .
\end{equation}
Our next step is to use this result to find some recursion relations for our diagrams. First, following equations (\ref{2.3.2}) and (\ref{2.5}) the $D_3$ diagram in terms of one node diagrams is given by,
\begin{equation}\label{2.14}
D_3
(\Lambda,Q_{s})=
   \frac{1}{2} \sum_{s_{2}} \frac{ N_{1}(b,0,Q_{2})N_{2}(b,1,Q_{2})+s_{2}N_{1}(b,0,\bar{Q}_{2})N_{2}(b,1,\bar{Q}_{2}) } { (s_{2}q^{1/2})_{\infty} }
  \end{equation}  
where we have defined $\bar{Q}_{2}=Q_{2}+1 \mod{2}$ and summed over $s_1=\pm1$ so that linear terms in $s_1$  vanish.
In diagrammatic language this relation is given by,
\begin{equation}\label{2.15}
\includegraphics[width=0.7\linewidth,keepaspectratio=true]{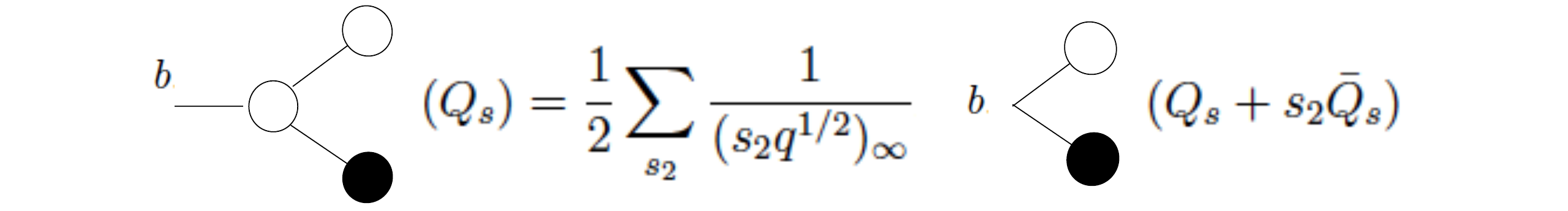}
\end{equation}
such that $Q_s=Q_{\pm}$ is a vector of length equal to the number of the associated diagram nodes, furthermore $Q_{+}=\bar{Q}_{-}$. Finally, if we regard the diagram as part of an $SO(2r)$ diagram with $\Lambda=\omega_r$ and $Q=Q_{\pm}$ we find these diagrams follow the same recursion relation as $\chi_{t,\pm}$ respectively.  \\
To prove our conjecture, for $\Lambda=\omega_r$ and $Q=Q_{\pm}$, we still need to prove it for $r=2$ diagrams which are easily calculated using eq. (\ref{2.5}),
\begin{equation}\label{2.16}
\includegraphics[width=0.18\linewidth,keepaspectratio=true]{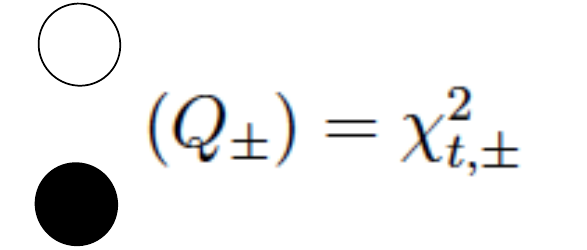}
\end{equation}  
We have thus proven the conjectured identities eq. (\ref{2.3.1}). It should be clear that this proof has a some what more general feature. The diagrammatic reduction relation eq. (\ref{2.15}) is a broader mathematical statement and can be applied to any sum which contains the $D_3$ diagram. This entails an infinite number of equivalent sums identities which we can write as, 
\begin{eqnarray}\label{2.22}
\text{\includegraphics[width=0.7\linewidth,keepaspectratio=true]{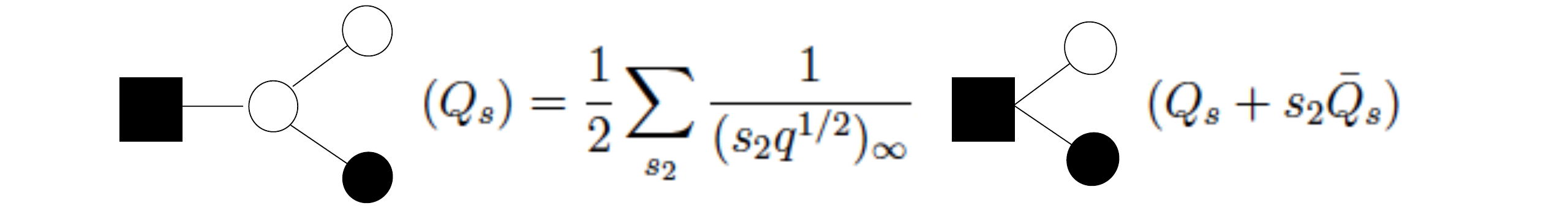}}
\end{eqnarray}  
where the black box stands for any sum(not necessarily diagrams). Furthermore, exact sums can be written by taking the opposite route, i.e start with some solvable diagram containing the $SO(4)$ tail and develop some $SO(2r)$ "tail".  
We also note that the $D_3$ diagram possess a $Q$ symmetry which entails additional non trivial identities.\\ 
Finally, to complete the proof of our conjecture for $SO(2r)$ diagrams of $\Lambda=\omega_i$, for $i=r-1,r$, we should consider other values of $Q$. This is the subject of our next section, in which will show that the $D_r(\Lambda,Q)$ diagrams for any $Q$ turn out to be equivalent to one of the cases studied in this section.

\subsection{\large{Twisted sector equivalent $\boldsymbol{Q}$ representations}  }

To show that the diagrams corresponding to $\Lambda=\omega_r$ and any  $Q$ are equivalent to one of the cases studied in the last section we study the equivalent $Q$ representations of  $Q_\pm$. First, denote the weight lattice of $SO(2r)$ by $P$. Next, using equation (\ref{1.9}) we find $\lambda$ mod $2P$ associated with $Q_\pm$ denoted by $\lambda_\pm$,
\begin{equation}\label{2.23}
\lambda_\pm=\Lambda-Q_{\pm}=\Lambda \mod 2P,
\end{equation} 
note that $Q_{\pm}=0 \mod2P$.
As a simple example, consider the rank $r=4$ equivalent $Q$ representation of highest weight $\lambda_{+}$ denoted $\lambda_{+}^r$. Starting from $\lambda_+=(0,0,0,1)$ we can subtract $\alpha_4$, which is read off the Cartan matrix,
\begin{eqnarray}\label{2.24}
\nonumber \lambda_+ - \alpha_4=(0,1,0,1) \mod 2P, \\
Q_+ - \alpha_4=\alpha_4 \mod 2M.
\end{eqnarray} 
Since we do not allow subtracting the same simple root twice in succession we can only proceed by subtracting $\alpha_2$, we find  
\begin{eqnarray}\label{2.25}
\nonumber \lambda_+ - \alpha_4-\alpha_2=(1,1,1,0) \mod 2P, \\
Q_+ - \alpha_4 - \alpha_2=\alpha_4 + \alpha_2 \mod 2M.
\end{eqnarray} 
To proceed we need to subtract $\alpha_1$ or $\alpha_3$ and so forth in a similar fashion. Finally, we find the $\lambda^4_+$  \\
\begin{minipage}{0.7\textwidth}
\vspace{0.05in}
representation is given by the figure on the right. There are a few features of the $\lambda^4_+$ representation which are worth mentioning. First, note that $\lambda_-= \lambda_+$, thus we find $\lambda^4_+=\lambda^4_-$, in the sense that they include the same weights. As we saw in the previous section, the diagrams corresponding to $Q_\pm$ are not equivalent. We thus conclude that although the $\lambda^4_\pm$ representations include the same weights their corresponding diagrams are necessarily different. Second, the dimension of $\lambda^4_+$ is given by $2^3$. These weights correspond to $2^3$ different $Q$ values for which the diagrams are equivalent. Together with the $2^3$ values of $Q$ corresponding to $\lambda^4_-$ we find the diagrams corresponding to all values of $Q$ are equivalent to one of the diagrams $D_4(\omega_4,Q_\pm)$. Third, if we do not allow subtraction of $\alpha_1$ we find the following decomposition  $\lambda^4_+=\lambda^3_+ \oplus \bar{\lambda}^3_+$. Where bar denotes exchanging the last two Dynkin labels, i.e $\bar{\lambda}_+=(0,1,0)$. Finally, due to this mapping of $\bar{\lambda}^3_+$ and $\lambda^3_+$, $\text{dim}\bar{\lambda}^3_+=\text{dim}\lambda^3_+$. 
\vspace{0.05in}
\end{minipage}
{\begin{minipage}{0.3\textwidth}
\begin{center}
\footnotesize{$\lambda_+^4$ equivalent $Q$ representation.}
\includegraphics[width=0.57\linewidth,keepaspectratio=true]{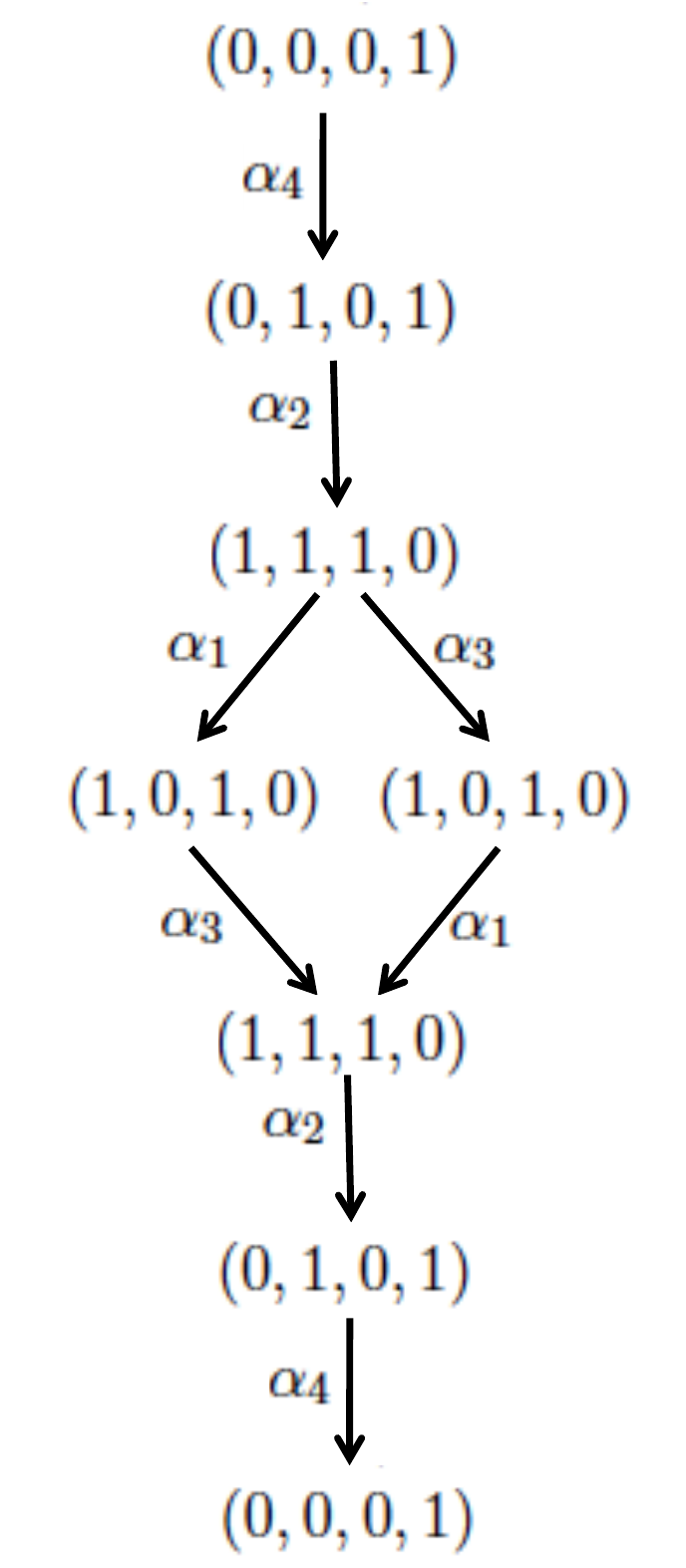}
\end{center}
\end{minipage} 
To show that for any rank $r$ all cases of $Q$ are equivalent to either $Q_+$ or $Q_-$ we prove these features. First, at any rank $\lambda_+=\lambda_-$ it follows that,
\begin{eqnarray}\label{2.27}
\lambda^r_+=\lambda^r_-,
\end{eqnarray}
in the sense that they include the same weights.
As for the $r=4$ case, we know from the previous section that diagrams corresponding to $Q_\pm$ are not equivalent and so diagrams corresponding to these representations are necessarily different. Thus, the dimension of these representations is equal and bounded from above by $2^{r-1}$. Next, clearly the decomposition of $\lambda_+^r$ into representations of rank $r-1$ includes $\lambda^{r-1}_+$. To show that it also includes the $\bar{\lambda}^{r-1}_+$ representation consider the following sequence appearing in the $\lambda_+^r$ representation,
\begin{eqnarray}\label{2.28}
\lambda_+,\lambda_+ - \alpha_r,\lambda_+ - \alpha_r - \alpha_{r-2},..., \lambda_+ - \sum^{r}_{ i=1 \atop i \neq r-1}\alpha_{i}.
\end{eqnarray}
The last weight of this sequence is given by,    
\begin{eqnarray}\label{2.28}
 \lambda_+ - \sum^{r}_{ i=1 \atop i \neq r-1}\alpha_{i}=(1,0,...,1,0).
\end{eqnarray}
Which shows that if we do not allow subtractions of $\alpha_1$ we find the following decomposition,
\begin{eqnarray}\label{2.29}
\lambda^r_+=\lambda^{r-1}_+ \oplus \bar{\lambda}^{r-1}_+ \oplus ....
\end{eqnarray}
where the dots stand for any other representation that might appear in the decomposition. That the two representations $\lambda^{r-1}_+$ and $\bar{\lambda}^{r-1}_+$ are equivalent under the exchange of the last two Dynkin labels follows from Lie algebra representations, thus the dimension of these representations is equal and we find,
\begin{eqnarray}\label{2.30}
\text{dim}\lambda^r_+=2\text{dim}\lambda^{r-1}_+ + ....
\end{eqnarray}
which are recursive relations for $\text{dim}\lambda^r_+$.
Assuming  $\text{dim}\lambda^{r-1}_+=2^{r-2}$,
\begin{eqnarray}\label{2.31}
\text{dim}\lambda^r_+=2^{r-1} 
\end{eqnarray}
where we have eliminated the dots since the dim$\lambda^r_+$ is bounded by $2^{r-1}$. Finally, using our example for $r=4$ as an initial condition we find,  
\begin{eqnarray}\label{2.32}
\text{dim}\lambda^r_+ + \text{dim}\lambda^r_-=2^{r}. 
\end{eqnarray}
Since each weight in these representations corresponds to some unique $Q$ we conclude that the $D_r(\omega_r,Q)$ diagrams for any root $Q$ are equal to either $D_r(\omega_r,Q_+)$ or $D_r(\omega_r,Q_-)$. The identification is easily achieved by using the invariance of $\lambda_{\pm}^2$ under the Weyl group. For a general $Q$ define $\lambda_Q=\Lambda-Q$ then,
\begin{eqnarray}\label{2.33}
\lambda_{+}^2/2-\lambda_Q^2/2=0,1 \mod2
\end{eqnarray}
where its $0$ if $Q \in \lambda^r_+$ and $1$ if $Q \in \lambda^r_-$. To state our result, consider,
\begin{equation}\label{2.34}
\includegraphics[width=0.41\linewidth,keepaspectratio=true]{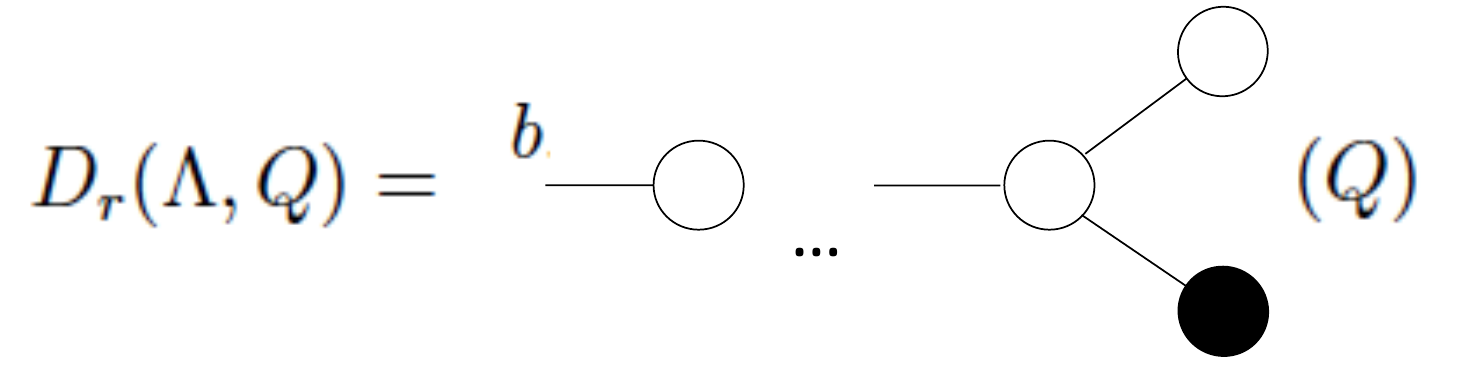}
\end{equation}
for a general $Q$ and $\Lambda=b\omega_1+\omega_r$. Clearly, the diagrams corresponding to our current discussion are given by  $D_r(\omega_r,Q)$ i.e $b=0$. However, recall that $\lambda_{\pm}$ are only defined $\mod2P$ thus the results above are viable for any even $b$. Additionally, by considering $\lambda_{\pm}=\Lambda-Q \mod2P$ for odd $b$ it is easily verified that eq. (\ref{2.31}) holds for any $b$. Finally, recalling the diagrammatic recursion relations of eq. (\ref{2.15}) we find the following identity, 
\begin{equation}\label{2.35}
D_r(\Lambda,Q)=
  \frac{1}{4}  \sum_{s,s_{1}}  s^{\frac{1}{2}Q^2+Q\Lambda} \frac{  (-s_{1}q^{1/2-b/2})_{\infty} (-s_{1}sq^{-b/2})_{\infty} } {(sq^{1/2})^{r-2}_{\infty}}
 \end{equation}
for any positive $b$ integer. The case of $b=0$ reproduces our conjecture for the twisted sector characters and concludes the proof of our conjecture for the twisted sector \footnote{That the character corresponding to some $Q$ has the "correct" dimension follows trivially from the invariance of the scalar product under the Weyl group.}.

\section{\large{The zero momenta diagrams}  }

The identities for the zero momenta characters were discussed in our paper \cite{gengep} and are given by our conjecture eq. (\ref{-2.4}). Specifically, calculating the dimensions for $\Lambda=0$ and $\lambda_{+/-}=0,\alpha_{r-1}+\alpha_r$ we conjecture the following identities corresponding to the zero momenta fields characters of eq. (\ref{2.1}),
\begin{equation}\label{3.1}
 \includegraphics[width=0.37\linewidth,keepaspectratio=true]{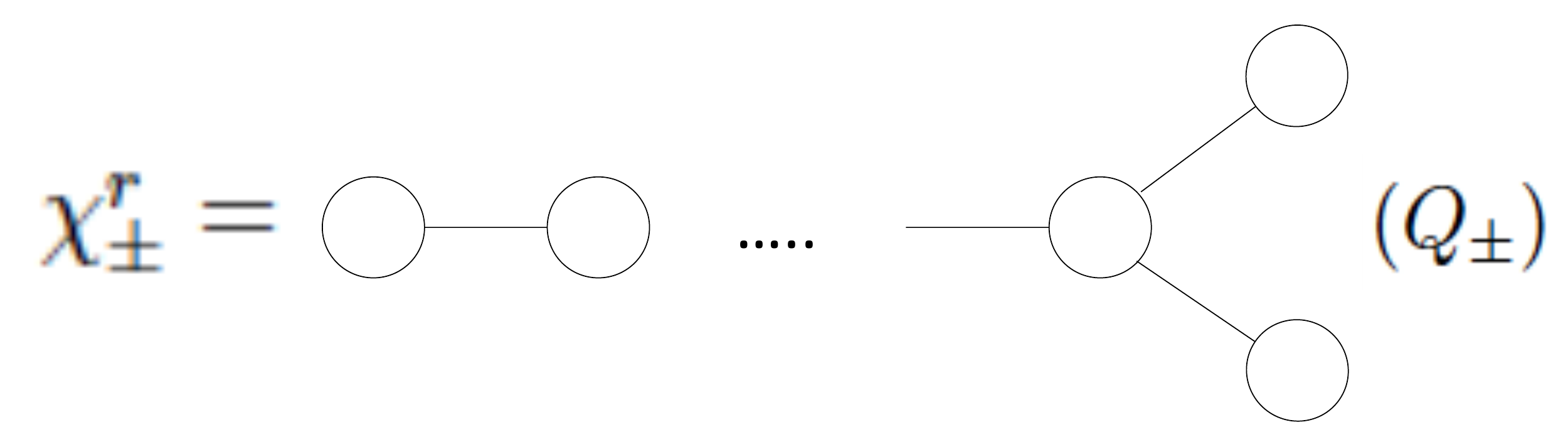}
 \end{equation}
where $Q_{+}=(0,...,0)$ and $Q_{-}=(0,...,1,1)$ are $SO(2r)$ roots. However, unlike the twisted sector, our expressions for these characters, eq. (\ref{2.1}), do not follow a nice recursion relation. The first part of the solution to this problem is defining the following combinations,
\begin{equation}\label{3.2}
\psi^r_{\pm}=\chi^r_+ \pm \chi^r _-.
 \end{equation}
Motivated by our proof for the twisted sector, we note that $\psi^r_-=(-q)^{1-r}_{\infty}$ follows a similar recursion relation to that of the twisted sector(eq. \ref{2.4}), 
\begin{equation}\label{3.3}
\psi^{r}_{-}= (-q)_{\infty}^{-1}\psi^{r-1}_{-}.
\end{equation}
This recursion relation implies that we can continue in a similar fashion to the twisted sector. Indeed, this is expected as $\psi_-$ is just the contribution of the twisted sector in the time direction. Consider the corresponding $D_3(\Lambda,Q)$ diagram,
\begin{equation}\label{3.4}
 \includegraphics[width=0.55\linewidth,keepaspectratio=true]{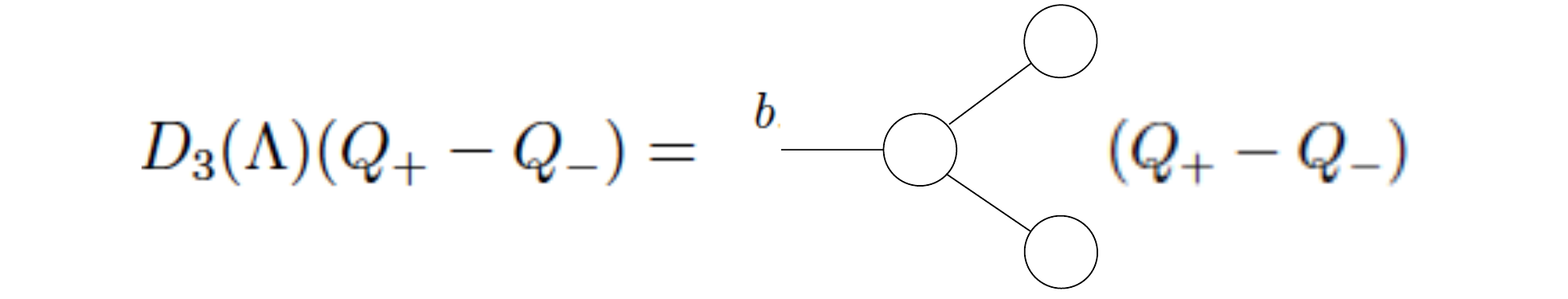}
 \end{equation}
where $\Lambda=b\omega_1$ and we subtract the two diagrams corresponding to $Q_+=(0,0,0)$ and $Q_-=(0,1,1)$. These diagrams are solved by first solving the second and third nodes and setting $Q_{2}=Q_3$,
\begin{equation}\label{3.5}
N_{2}(b_1,0,Q_3)N_{3}(b_1,0,Q_3)=
\frac{1}{4}\sum_{s_{2},s_{3}}     
     \frac{(-s_{2})^{Q_{3}} (s_{2}q^{1/2})_{\infty} }   { (s_{2}q^{1/2})_{-a_{1}} }
    \frac{(-s_{3})^{Q_{3}} (s_{3}q^{1/2})_{\infty} }   { (s_{3}q^{1/2})_{-a_{1}} } .
\end{equation}
Subtracting the two diagrams corresponding to $Q_3=0$ and $Q_3=1$ we find that the contribution of $s_{2} = s_3$ vanishes, thus we set $s_{2}=-s_3$. Moreover, using the Pochhammer identities eq. (\ref{2.10}) we find,
\begin{equation}\label{3.6}
D_3
(b\omega_1)(Q_+-Q_-)=
A \sum_{a_{1}=0}^{\infty}   \frac {q^{a_{1}(a_{1}-b)} (-1)^{a_{1}} }  { 
   (q)_{a_{1}}     (-q)_{a_{1}} },
\end{equation}
where $A$ stands for the $a_{1}$ independent terms,
\begin{equation}\label{3.7}
A=\frac{1}{2}\sum_{s_{3}}  
    (s_{3}q^{1/2})_{\infty} 
    (-s_{3}q^{1/2})_{\infty} .
\end{equation}
To solve the sum over $a_{1}$ we invoke the Euler identity eq. (\ref{1.4}) at $q \rightarrow q^2$ and $z=q^{1-b}$.
Performing the summation over $s_{3}$ and using $(-q)_{\infty}(-q^{1/2})_{\infty}(q^{1/2})_{\infty}=1$ the diagram is given by,
\begin{equation}\label{3.8}
D_3(b\omega_1)(Q_+-Q_-)=
(-q)^{-1}_{\infty}  (-q^{1/2-b/2})_{\infty}  (q^{1/2-b/2})_{\infty}.
\end{equation}
To identify the diagrammatic relation implied by this solution write it in terms of the one node diagrams using eqs. (\ref{2.5}) and (\ref{2.3.2}),
\begin{equation}\label{3.9}
D_3
(b\omega_1)(Q_+-Q_-)=
(-q)^{-1}_{\infty}   \sum_{Q}  (-1)^Q N_{2}(b,0,Q) N_{1}(b,0,Q)  
\end{equation}  
and draw the corresponding diagrams,
\begin{equation}\label{3.10}
\includegraphics[width=0.66\linewidth,keepaspectratio=true]{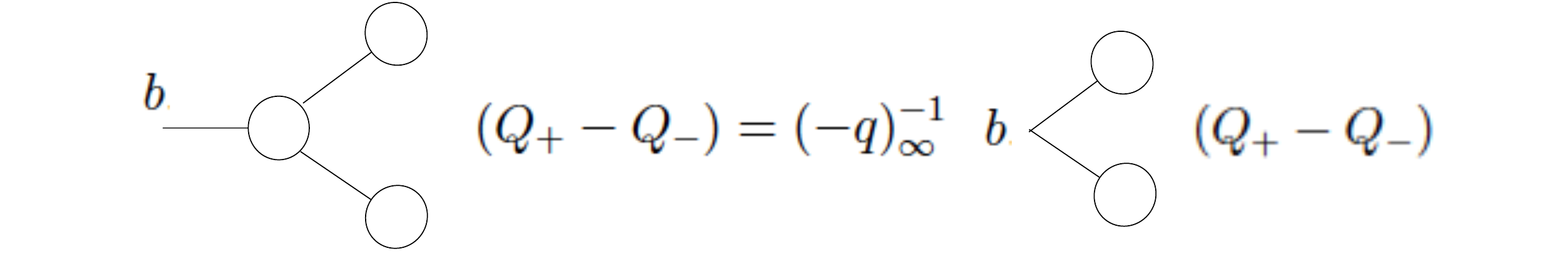}
\end{equation}
 which as expected is the same recursion relation as for $\psi_-$. Clearly, this means that any $SO(2r)$ diagrams combination of $Q_+-Q_-$ satisfies the same recursion relation as $\psi_-$.  Recalling we have already established the conjectured identities for rank $r=2,3$ diagrams in section ($4$), see eqs. (\ref{2.3.3}) and (\ref{2.3.8}) respectively.  We find the expected identity, equation (\ref{2.3.9}) of section ($4$), corresponding to the contribution of the  time direction twisted sector of $SO(2r)$. Actually, using the recursion relations we find the following identity,
\begin{equation}\label{3.14}
 \includegraphics[width=0.75\linewidth,keepaspectratio=true]{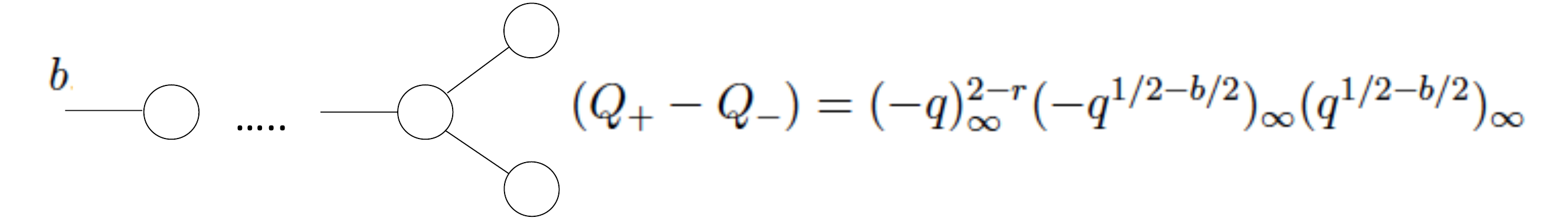}
 \end{equation}
here the diagram has $r$ nodes while $Q_+=(0,...,0)$ and $Q_-=(0,...,1,1)$ are rank $r$ roots. Let us note that,  as in the case of the twisted sector 
the diagrammatic recursion relation eq. (\ref{3.10}) is a general mathematical statement. As a simple example, consider any solvable sum containing the $SO(4)$ time direction twisted sector tail,
\begin{equation}\label{3.15}
 \includegraphics[width=0.45\linewidth,keepaspectratio=true]{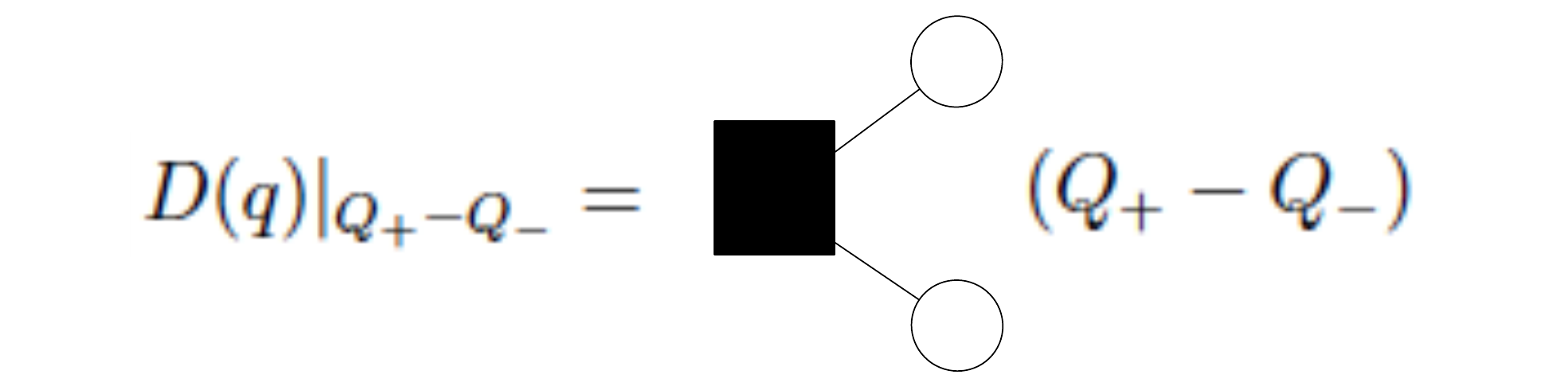}
 \end{equation}
where $Q_{\pm}$ are vectors of integers mod $2$ of length two. Following the diagrammatic recursion relations, eq. (\ref{3.10}), we can immediately write down the identity corresponding to a general $SO(2r)$ tail, 
\begin{equation}\label{3.15}
 \includegraphics[width=0.64\linewidth,keepaspectratio=true]{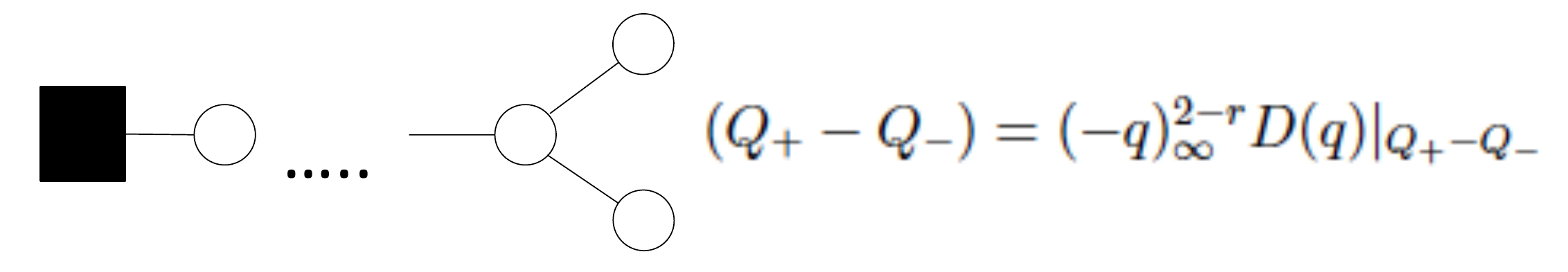}
 \end{equation}
where $Q_{\pm}$ are vectors of length equal to that of their corresponding diagram given by $Q_+=(0,...,0)$ and $Q_-=(0,...,1,1)$. We will come back to these identities in the proceeding sections as they provide us with some interesting possibilities of generalizing our conjecture. Even more fascinating is the possibility to generalize the classic works of Euler, Heine and Cauchy which have all played a part in our discussion. \\
Returning to the problem at hand, having proven the identities corresponding to the time direction twisted sector contribution we are left with the contribution of $r-1$ free bosons on the $SU(r)$ level $2$ lattice,
\begin{equation}\label{3.15}
\psi^r_+=\Theta^r_{0}(q) (q)_{\infty}^{-r+1}
 \end{equation}
where the level $2$ classical theta function $\Theta^r_{\lambda}(q)$ was defined in eq. (\ref{2.0}). Our first obstacle is writing a recursion relation for $\psi^r_+$. Let us take a closer look at $\Theta_{\lambda}(q)$, 
\begin{eqnarray}\label{3.16}
\Theta^r_{\lambda}(q)=\sum_{\mu\in M_{SU(r)}+\lambda/2} q^{u^2}=\sum_{\{n_i\} =-\infty}^{\infty} q^{(n + \lambda/2)^2}
\end{eqnarray}
here $M_{SU(r)}$ denotes the root lattice of $SU(r)$, $\lambda=\sum \lambda_i\omega_i$ is a weight of $SU(r)$, $n=\sum n_i\alpha_i$ is a root of $SU(r)$ and the summation is over all integer $n_i$'s where $i=1,..,r-1$. To proceed, note that any weight of $SU(r)$ can be written as:
\begin{eqnarray}\label{3.17}
\lambda=n-a\omega_{r-1},
\end{eqnarray}
with some integer $a$ and $n \in M_{SU(r)}$. Actually, the level $2$ theta function is symmetric under any even root translation of $\lambda$,
\begin{eqnarray}\label{3.18}
\Theta^r_{\lambda}(q)=\Theta^r_{\lambda+2n}(q),
\end{eqnarray}
as this amounts to shifting the summation variables by $n_i$.  Thus, it is enough to consider only $\lambda$ mod $2M_{SU(r)}$  as all other cases are equivalent, 
\begin{eqnarray}\label{3.19}
\lambda=Q-a\omega_{r-1}   \mod 2M_{SU(r)}
\end{eqnarray}
where $Q=\sum Q_{i}\alpha_i$ is a root of $SU(r)$ such that $Q_{i}=0,1$. Let us use this decomposition to rewrite $\Theta^r_{\lambda}(\tau)$ as,
\begin{eqnarray}\label{3.20}
\Theta^r_{Q-a\omega_{r-1}}(\tau)=\sum_{\{n_i\} =-\infty \atop n_i \in Z + Q_{i}/2}^{\infty} q^{(n - a\omega_{r-1}/2)^2}.
\end{eqnarray}
To write this sum explicitly note that here $n$ is a root of $SU(r)$ with fractional coefficients while $\omega_{r-1}$ is a fundamental weight. Denoting the Cartan matrix by $A_{ij}$ we find,
\begin{eqnarray}\label{3.21}
(n - a\omega_{r-1}/2)^2=n_i A_{i,j} n_j-n_{r-1}a+a^2(r-1)/4r=\sum^{r-1}_{i=1}(2n^2_i-2n_{i}n_{i+1})+\Delta_{2n_r}.
\end{eqnarray}
Here to make the recursion apparent we set $a=2n_r$ as well as $a=Q_r$ mod $2$, so that  $n_r \in Z +Q_r/2$ and $\Delta_{2n_r}=n_{r}^2(r-1)/r$. Let us now define,
\begin{eqnarray}\label{3.18a}
\psi^r_{2n_r,{Q}}(q)=(q)^{-r+1}_{\infty}  q^{-\Delta_{2n_r}}   \Theta^r_{Q-2n_r\omega_{r-1}}(q)=(q)^{-r+1}_{\infty} \sum_{\{n_i\} =-\infty \atop n_i \in Z + Q_{i}/2}^{\infty} q^{n_i A_{i,j} n_j-2n_{r-1}n_r}.
\end{eqnarray}
Indeed, $\psi^r_{2n_r,Q}$ follows a trivial recursion relation,
\begin{eqnarray}\label{3.19b}
\psi^r_{2n_r,Q}(q)=(q)^{-1}_{\infty}\sum_{n_{r-1}=-\infty \atop n_{r-1} \in Z + Q_{r-1}/2}^{\infty}q^{2n_{r-1}^2-2n_{r-1}n_r}\psi^{r-1}_{2n_{r-1},Q}(q),
\end{eqnarray}
where $\psi^1_{2n_1,Q}=1$. Furthermore, note that here the $SU(l)$ root associated with $\psi^l_{2n_l,Q}$ is given by $Q=(Q_1,...,Q_{l-1})$.
Clearly, by finding some diagrammatic expression for $\psi^r_{2n_r,Q}(q)$ we can verify our conjecture as,
\begin{equation}\label{3.20c}
\psi^r_{0,0}(q)  =\psi^r_+(q).
 \end{equation}
To gain some intuition regarding the diagrams corresponding to $\psi^r_{2n_r,Q}$ examine the simplest case of $r=2$ which, via a change of variables $2n_1=m_1$, can be written  as:
\begin{equation}\label{3.21d}
\psi^2_{2n_2,Q_1}=(q)^{-1}_{\infty}\sum_{n_{1}=-\infty \atop n_{1} \in Z + Q_{1}/2}^{\infty}q^{2n_{1}^2-2n_{1}n_2}=\frac{1}{2}(q)^{-1}_{\infty}\sum_{m_{1}=-\infty }^{\infty}q^{m_{1}^2/2-m_{1}n_2}(1+(-1)^{m_1+Q_1}).
 \end{equation}
The sums appearing here are solved using the Jacobi identity, eq. (\ref{2.3.10}), given in section ($4$). Specifically, solving the sum over $m_1$ we find that $\psi^2_{2n_2,Q_1}$ is given by,
\begin{equation}\label{3.22}
\psi^2_{2n_2,Q_1}=\frac{1}{2}\sum_{s=\pm1}s^{Q_1}(-sq^{1/2-n_2})_{\infty}(-sq^{1/2+n_2})_{\infty}.
 \end{equation}
This expression is quite similar to our expression for $2$ non connected nodes eq. (\ref{2.5}). More specifically, the sign of $n_2$ in the second Pochhammer symbol is wrong. However, we can  encode this information by introducing a new diagrammatic rule,
\begin{enumerate}[label=\roman{*}., ref=(\roman{*})]
\item \text{for each dashed line connecting $b_i$ and $b_j$ } $=q^{b_ib_j/2}$.   
\end{enumerate}
We can now draw the diagram corresponding to $\psi^2_{2n_2,Q_1}$,
\begin{equation}\label{3.23}
 \includegraphics[width=0.68\linewidth,keepaspectratio=true]{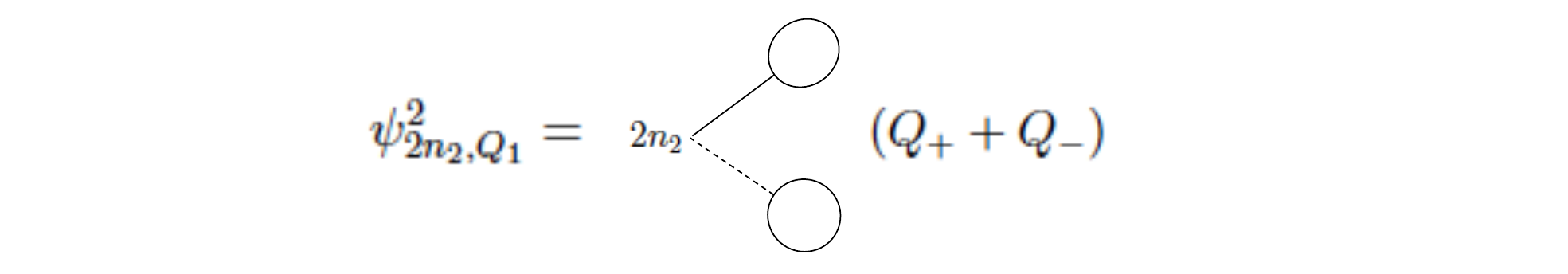}
 \end{equation}
where we defined the $SO(4)$ roots $Q_+=(Q_1,0)$ and $Q_-=(1-Q_1,1)$. Note, that indeed for $n_2=0$, this diagram corresponds to the $SO(4)$ diagram so that this proves our conjecture for $r=2$. To write down a diagrammatic recursion relation we need to find some diagrammatic expression for $\psi^3_{2n_3,Q}$. Such an expression, for $n_3=0$, was found as an example in section ($4$). Indeed, setting $z=n_3$ and manipulating the first Ramanujan identity, eq. (\ref{2.3.10.1}), in a similar fashion one can find the diagrammatic interpretation of $\psi^3_{2n_3,Q}$ for a general $n_3$. Actually, the expression for a general $n_3$ can be deduced, in a simpler way, via the following diagrammatic argument. The diagram corresponding to $\psi^3_{2n_3,Q}$ should be equal to the $SO(6)$ diagram when taking $n_3=0$. Moreover, we should be able to construct $\psi^4_{2n_4,Q}$ from this diagram with an external line connected to the first node. These diagrammatic arguments lead to the following conjectured identities,
\begin{eqnarray}\label{3.24}
\includegraphics[width=0.63\linewidth,keepaspectratio=true]{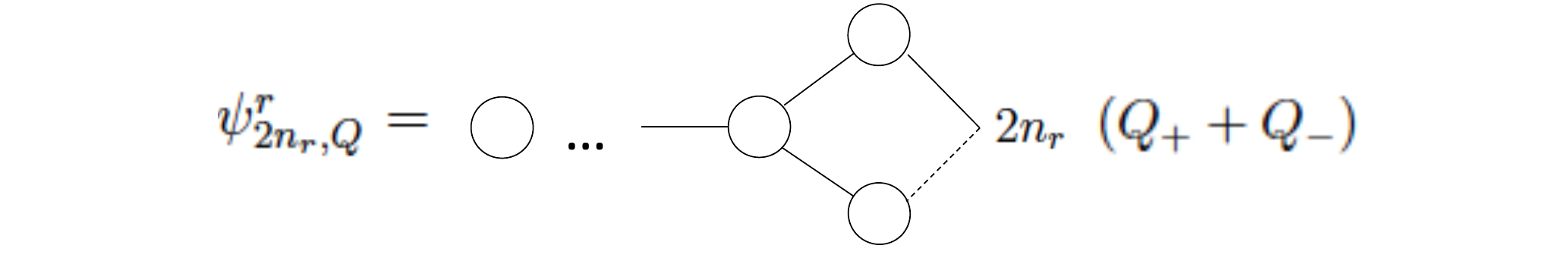}
\end{eqnarray}
where note that while $Q$ is a root of $SU(r)$ we define the $SO(2r)$ roots $Q_+=(Q_1,...,Q_{r-1},0)$ and $Q_-=(Q_1,...,1-Q_{r-1},1)$.
It should be  clear that these identities are a generalization of our conjecture motivated by the diagrammatic interpretation of the Jacobi and Ramanujan identities, eqs. (\ref{2.3.10}) and (\ref{2.3.10.1}), for a general $z$. To prove these identities let us examine the following 
$D_3(\Lambda,Q)$ diagram,
\begin{eqnarray}\label{3.25}
\includegraphics[width=0.6\linewidth,keepaspectratio=true]{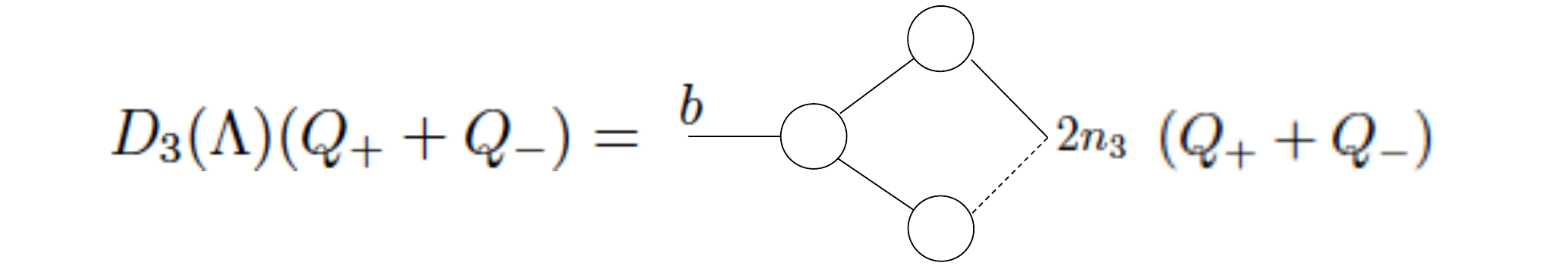}
\end{eqnarray}
Where $\Lambda=b\omega_1+2n_3(\omega_{r-1}-\omega_{r})$, $Q_+=(Q_1,Q_2,0)$ and $Q_-=(Q_1,1-Q_2,1)$. We can write down the sum corresponding to this diagram,
\begin{eqnarray}\label{3.26}
D_3(\Lambda)(Q_++Q_-)= \sum_{Q=Q_{\pm}}\sum^{\infty}_{\{b_i\}=0 \atop b_i=Q_i \mod2}\frac{q^{(b_1^2+b_2^2+b_3^2-b_1b_2-b_1b_3-b_1b-2b_2n_3+2b_3n_3)/2}} {(q)_{b_1}(q)_{b_2}(q)_{b_3}}
\end{eqnarray}
where the sum is over $b_1,b_2$ and $b_3$. Up to now our strategy for solving diagrams was to start by solving the last two nodes. However, this approach is not fruitful in the case at hand. Instead, using the one node solution eq. (\ref{2.8}) for the first node,
\begin{eqnarray}\label{3.27}
D_3= \frac{1}{2}\sum_{s_1=\pm1}\sum_{Q=Q_{\pm}}\sum^{\infty}_{b_2,b_3=0 \atop b_i=Q_i \mod2}\frac{q^{(b_2^2+b_3^2-2b_2n_3+2b_3n_3)/2}} {(q)_{b_2}(q)_{b_3}}s_1^{Q_1}(-s_1q^{(1-b_2-b_3-b)/2})_{\infty},
\end{eqnarray}
where we abbreviate $D_3=D_3(\Lambda)(Q_++Q_-)$.
Lets take a closer look on the restriction over $b_2$ and $b_3$. Since we are summing over the same diagram with different $Q$ roots we can simply sum over the $Q$ dependent factors,
\begin{eqnarray}\label{3.28}
\frac{1}{4}\sum_{Q=Q_{\pm}}(1+(-1)^{b_2+Q_2})(1+(-1)^{b_3+Q_3})=
\frac{1}{2}(1+(-1)^{b_2+b_3+Q_2})
\end{eqnarray}
Implementing  this result, 
\begin{eqnarray}\label{3.29}
D_3= \sum_{s_1=\pm1}\sum^{\infty}_{b_2,b_3=0}\frac{q^{(b_2^2+b_3^2-2b_2n_3+2b_3n_3)/2}} {4(q)_{b_2}(q)_{b_3}}s_1^{Q_1}(-s_1q^{(1-b_2-b_3-b)/2})_{\infty}(1+(-1)^{b_2+b_3+Q_2}).
\end{eqnarray}
Before we proceed, let us take a moment to reflect about the recursion relation we are looking for. 
To prove the conjectured identity  eq. (\ref{3.24}) we should find that $D_3$ follows the same recursion relation as $\psi^r_{2n_3}$ eq. (\ref{3.19b}). 
In contrary to the recursion relations we examined up to now, the $D_3$ recursion relation would contain the sum over an infinite number of diagrams. Furthermore, this summation would range from minus to plus infinity and not necessarily over integers. While the summations appearing in our diagram are taken over integers which range only from zero to infinity. These complications are actually a blessing in disguise as they imply that to proceed we should look for change of variables that will produce said summation. Actually, this change of variable is quite natural from the diagrammatic point of view. Recall that the diagrams corresponding to $\psi^r_{\pm}$  are the same however with a different combination of $Q$ roots. This change of $Q$ roots combination should lead to the different recursion relations. Indeed, summing over $Q_{\pm}$ (eq. \ref{3.28}) restricts $b_2$ and $b_3$ to the same parity for $Q_2=0$ or the opposite parity for $Q_2=1$ which implies the change of variables,
 \begin{eqnarray}\label{3.29}
b_3=b_2+2n_2.
\end{eqnarray}
So that $n_2 \in Z+Q_2/2$ and the summation over $n_2$ now ranges from minus to plus infinity\footnote{One should note that the Pochhammer symbol limits the summation to the desired range.},
\begin{eqnarray}\label{3.30}
D_3= \frac{1}{2}\sum_{s_1=\pm1}\sum^{\infty}_{n_2=-\infty   \atop n_2 \in Z+Q_2/2}\sum^{\infty}_{b_2=0}\frac{q^{b_2^2+2b_2n_2+2n_2^2+2n_2n_3}} {(q)_{b_2}(q)_{b_2+2n_2}}s_1^{Q_1}(-s_1q^{(1-b)/2-b_2-n_2})_{\infty}.
\end{eqnarray}
To solve the sum over $b_2$ we first extract the $b_2$ dependance of the last Pochhammer,
\begin{eqnarray}\label{3.31a}
(-s_1q^{(1-b)/2-b_2-n_2})_{\infty}=\frac{(-s_1q^{(1-b)/2-n_2})_{\infty}}{(-s_1q^{(1-b)/2-n_2})_{-b_2}}.
\end{eqnarray}
The Pochhammer in the denominator can be manipulated using eq. (\ref{2.3.5}),
\begin{eqnarray}\label{3.31b}
(-s_1q^{(1-b)/2-n_2})_{-b_2}=\frac{s_1^{b_2}q^{b_2(b_2+b+2n_2)/2}}{(-s_1q^{(1+b)/2+n_2})_{b_2}}.
\end{eqnarray}
Next, using the definition of the Pochhammer symbol,
\begin{eqnarray}\label{3.32}
(q)_{b_2+2n_2}=(q)_{2n_2}(q^{1+2n_2})_{b_2}.
\end{eqnarray}
Replacing these in the $D_3$ diagram expression eq. (\ref{3.30}),
\begin{eqnarray}\label{3.33a}
\nonumber
D_3= A
\sum^{\infty}_{b_2=0}\frac{s_1^{b_2}q^{b_2(b_2-b+2n_2)/2}} {(q)_{b_2}(q^{1+2n_2})_{b_2}}(-s_1q^{(1+b)/2+n_2})_{b_2},
\end{eqnarray}
\begin{eqnarray}\label{3.33a}
A=\frac{1}{2}\sum_{s_1=\pm1}\sum^{\infty}_{n_2=-\infty   \atop n_2 \in Z+Q_2/2 }
\frac{q^{2n_2^2+2n_2n_3}}{(q)_{2n_2}}s_1^{Q_1}(-s_1q^{(1-b)/2-n_2})_{\infty}.
\end{eqnarray}
Finally, the sum over $b_2$ can now be solved using Heine's $q$ sum\cite{RRS},
\begin{equation}\label{3.33aa}
\frac{(c/a)_{\infty}}{(c)_{\infty}}=
 \sum_{n=0}^{\infty}   \frac {c^nq^{n(n-1)/2} (-1)^{n}(a)_{n} }  { 
 a^n  (q)_{n}     (c)_{n} }.
\end{equation}
To find $D_3$ is given by,
\begin{eqnarray}\label{3.33b}
D_3(\Lambda)(Q_++Q_-)= \sum^{\infty}_{n_2=-\infty    \atop n_2 \in Z+Q_2/2}
\frac{q^{2n_2^2-2n_2n_3}}{2(q)_{\infty}}\sum_{s_1=\pm1}s_1^{Q_1}(-s_1q^{(1-b)/2-n_2})_{\infty}
 (-s_1q^{(1-b)/2+n_2})_{\infty}
\end{eqnarray}
where we have used $(q)_{2n_2}(q^{1+2n_2})_{\infty}=(q)_{\infty}$ and changed $n_2 \rightarrow -n_2$. The diagrammatic interpretation of this relation is just,
\begin{eqnarray}\label{3.34a}
\includegraphics[width=0.88\linewidth,keepaspectratio=true]{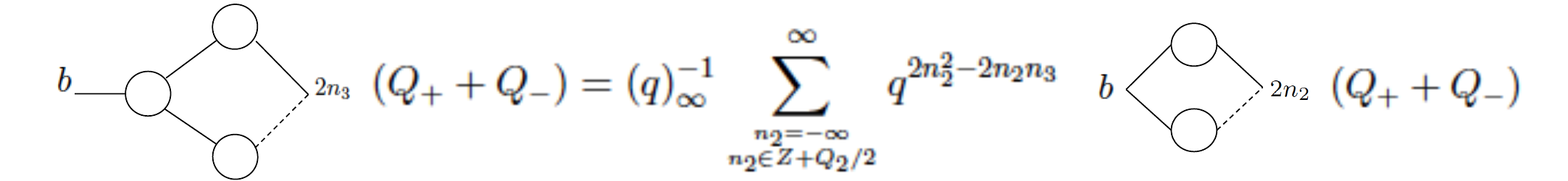}
\end{eqnarray}
which is exactly the expected relation. More specifically, the $D_3$ diagram is the tail of the conjectured diagrams of eq. (\ref{3.24}) . Thus, we find that these diagrams follow the same recursion relation as $\psi^r_{2n_r,Q}$. Since we already examined the $r=2$ case this concludes the proof of eq. (\ref{3.24}). Actually, just as for the twisted contributions, let us define the $r$ nodes diagram,
\begin{eqnarray}\label{3.35}
\includegraphics[width=0.65\linewidth,keepaspectratio=true]{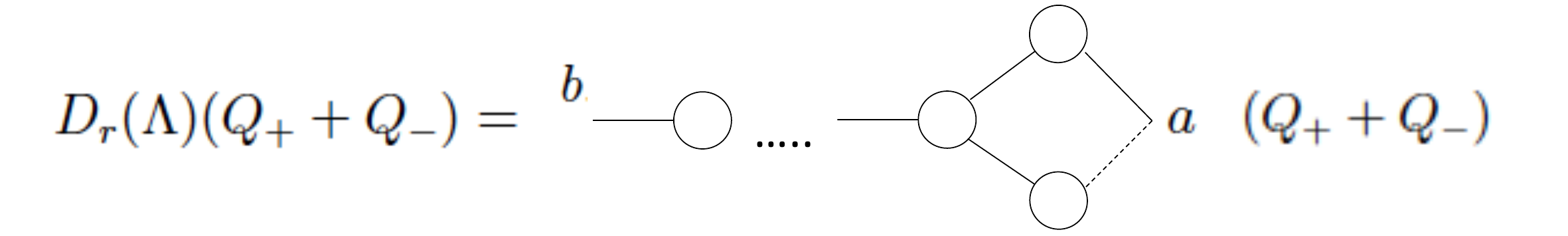}
\end{eqnarray}
where $\Lambda=b\omega_1+a(\omega_{r-1}-\omega_r)$.
Using the diagrammatic recursion relation, the Jacobi identity and redefining $2n_i \rightarrow n_i$ we find,
\begin{eqnarray}\label{3.36}
D_r(\Lambda)(Q_++Q_-)=\sum^{\infty}_{\{n_i\}=-\infty    \atop n_1 \in Z, n_i \in 2Z+Q_i}
\frac{q^{\frac{1}{4}n^2-\frac{1}{2}n_1b-\frac{1}{2}n_{r-1}a}}{2(q)^{r-1}_{\infty}}\sum_{s_1=\pm1}s_1^{Q_1+n_1} \frac{
 (-s_1q^{(1-b+n_2)/2})_{\infty}}{(-s_1q^{(1+b+n_2)/2})_{\infty}}.
\end{eqnarray}
Where $n=\sum n_i\alpha_i$ is a root vector of $SU(r)$, the summation is over $n_i$ for $i=1,...,r-1$ under the restriction $n_i=Q_i \mod2$ for $i=2,...,r-1$ and no restriction for $n_1$. Finally, it should be noted that the identity eq. (\ref{3.24}) is clearly given by the $b=0$ and $a=2n_r$ case. \\
Having found a diagrammatic expression for both $\psi_{-}$ (eq. \ref{3.14}) and $\psi_{+}$ (eq. \ref{3.24}),
\begin{eqnarray}\label{3.34b}
\includegraphics[width=0.73\linewidth,keepaspectratio=true]{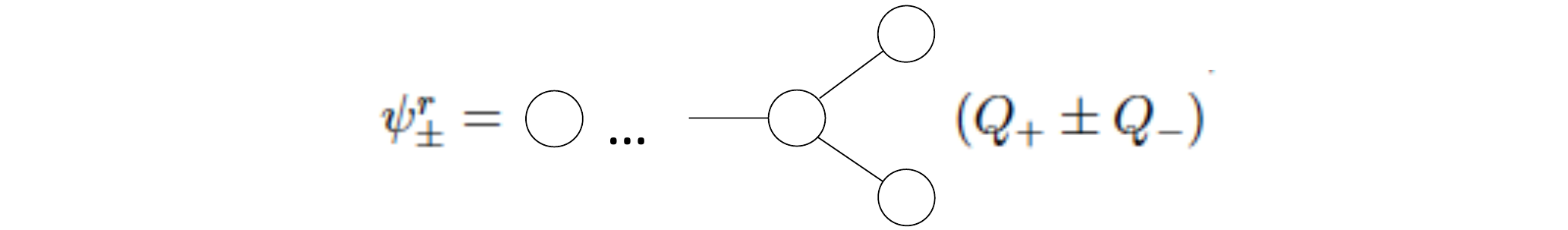}
\end{eqnarray}
where $Q_+=(0,...,0)$ and $Q_-=(0,...0,1,1)$.
Recall that the characters of the zero momenta fields are given by
\begin{eqnarray}\label{3.34c}
\chi^{r}_{\pm}=\frac{1}{2}(\psi^r_{+} \pm \psi^r_{-}),
\end{eqnarray}
from which trivially follow the conjectured identities for $\chi^r_{\pm}$ (eq. \ref{3.1}). This concludes the proof of the untwisted sector zero momenta characters. As in the case of the twisted sector, here we have considered the $\Lambda=0$ diagram for some chosen $Q$ values. It should be noted that to prove our conjecture we need to examine all values of $Q$. That none of the other values of $Q$ correspond to the zero momenta characters follows simply from the uniqueness of the decomposition (eq. \ref{3.19}) of $\lambda=0$ at $2n_r=0$.  
Instead, as it turns out, these correspond to the non-zero momenta characters, eq. (\ref{2.2}), which are the subject of our next section.

\section{\large{The nonzero momenta diagrams}  }

Following our conjecture, the nonzero momenta characters should also correspond to $SO(2r)$ diagrams.
Our objective in this section is to prove the identities which arise for the nonzero momenta characters via our conjecture.
The diagrams, we have yet to consider, correspond to either $\Lambda=0$ and $Q \neq 0,\alpha_{r-1}+\alpha_{r}$ or $\Lambda=\omega_1$. Let us continue our discussion from the previous section by first considering the $\Lambda=0$ diagrams. According to eq. (\ref{3.24}) sums of these diagrams are given simply by taking $n_r=0$,
\begin{eqnarray}\label{4.1}
\includegraphics[width=0.55\linewidth,keepaspectratio=true]{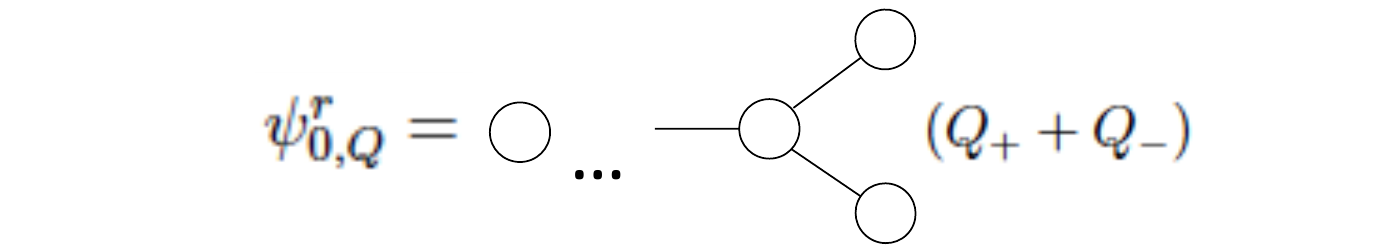}
\end{eqnarray}
where $Q=(Q_1,...,Q_{r-1}) \neq 0$ is a root of $SU(r)$ while $Q_+=(Q_1,...,Q_{r-1},0)$ and $Q_-=(Q_1,...,1-Q_{r-1},1)$ are $SO(2r)$ roots. To solve these diagrams we prove that for $Q \neq 0$ the diagrams appearing on the RHS are equivalent, 
\begin{equation}\label{4.2}
 \includegraphics[width=0.9\linewidth,keepaspectratio=true]{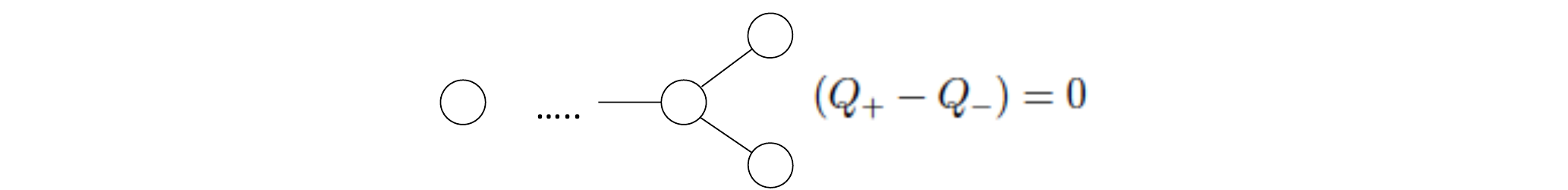}
 \end{equation}
Actually, since the $SO(2r)$ diagrams are symmetric under the exchange of the last two nodes, i.e $\Lambda_r \leftrightarrow \Lambda_{r-1}$ and $Q_r \leftrightarrow Q_{r-1}$. These identities follow trivially when $Q_{r-1}=1$ for which $Q_+=(Q_1,...,Q_{r-2},1,0) \leftrightarrow Q_-=(Q_1,...,Q_{r-2},0,1)$ under the mentioned symmetry. To prove these identities for $Q_{r-1}=0$ let us study the corresponding $Q_+=(Q_1,...,Q_{r-2},0,0)$ and $Q_-=(Q_1,...,Q_{r-2},1,1)$ equivalent Q representations. First, since $Q \neq 0$ we write without loss of generality, 
\begin{eqnarray}\label{4.3}
Q_{+}=(Q_1,...,Q_{i-1},1,0,...,0), \;\;\;\; Q_{-}=(Q_1,...,Q_{i-1},1,0,...,0,1,1),
\end{eqnarray}
for some $i=1,...,r-2$. Next, to calculate $\lambda_{+}$ recall eq. (\ref{1.9}),
\begin{eqnarray}\label{4.4}
\lambda_{+}=Q_{+} \mod 2P \Rightarrow \lambda_+=(\lambda_1,...,\lambda_{i},1,0,...,0)
\end{eqnarray}
where as usual $P$ is the $SO(2r)$ weight lattice and we have only specified the last $r-i$ Dynkin labels. To construct the $\lambda^+_r$ equivalent $Q$ representation note that $\lambda_{i+1}=1$, so we can start by subtracting $\alpha_{i+1}$ mod $2P$,
\begin{eqnarray}\label{4.5}
\lambda_{+}-\alpha_{i+1}=(\lambda_1,...,\lambda_{i}+1,1,1,0,...,0) \mod 2P,
\end{eqnarray}
to find an equivalent diagram with $Q=Q_+-\alpha_{i+1}$.
Next, note that $\lambda_{i+2}=1$, so we can continue to subtract $\alpha_{i+2}$, actually this can be repeated up to $\alpha_r$,
\begin{eqnarray}\label{4.6}
\lambda_{+}-\sum_{l=1}^{r-i}\alpha_{i+l}=(\lambda_1,...,\lambda_{i}+1,0,...,0,1,1,1) \mod 2P.
\end{eqnarray}
Finally, here $\lambda_{r-2}=1$  thus, we can now subtract $\alpha_{r-2}$. In a similar fashion, this can be repeated down to $\alpha_{i+1}$, 
\begin{eqnarray}\label{4.7}
\lambda_{+}-\alpha_{r-1}-\alpha_{r}=(\lambda_1,...,\lambda_{i},1,0,...,0) \mod 2P.
\end{eqnarray}
So this sequence of weights in the $\lambda^r_+$ representation just terminates at $\lambda_+$. The corresponding $Q$ root is given by subtracting these roots from $Q_+$,
\begin{eqnarray}\label{4.7}
Q_{+}-\alpha_{r-1}-\alpha_{r}=Q_- \mod 2M.
\end{eqnarray}
To conclude we find both $Q_{\pm} \in \lambda^{r}_+$, thus giving rise to equivalent diagrams which completes the proof of eq. (\ref{4.2}).  \\
However trivial, let us stress that the $\Lambda=0$ diagram for a general $SO(2r)$ root $Q=(Q_1,...,Q_r)\neq 0, \alpha_{r-1}+\alpha_r$ is now simply given by,
\begin{eqnarray}\label{4.8}
\includegraphics[width=0.55\linewidth,keepaspectratio=true]{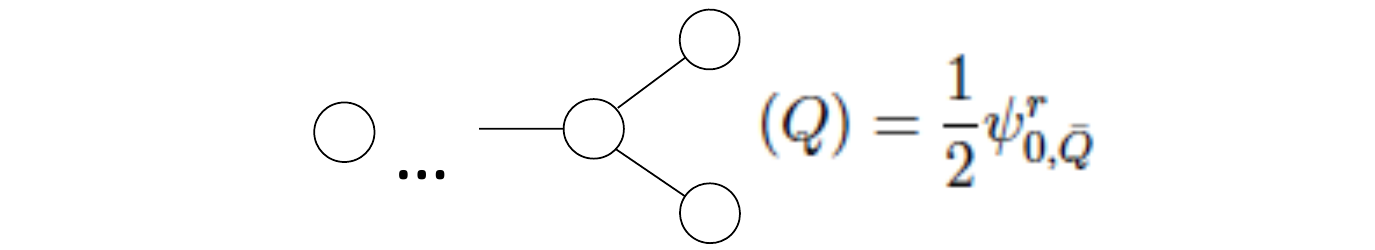}
\end{eqnarray}
where the $SU(r)$ root corresponding to $\psi^r_{0,\bar{Q}}$ is defined as $\bar{Q}_i=Q_i$ with $i=1,...,r-2$, while for $i=r-1$ we define $\bar{Q}_{r-1}=Q_r+Q_{r-1}$ mod $2$. The characters corresponding to $\psi^r_{0,\bar{Q}}$ follow from its definition eq. (\ref{3.18a}),   
\begin{eqnarray}\label{4.9}
\frac{1}{2}\psi^r_{0,{\bar{Q}}}(q)=\frac{1}{2}(q)^{-r+1}_{\infty}     \Theta^r_{\bar{Q}}(q)=\chi^{r}_{\bar{Q}},
\end{eqnarray}
where we have identified the nonzero momenta characters $\chi^r_{\lambda}$(see eq. \ref{2.2}) with $\lambda=\bar{Q}$, i.e weights which belong to the root lattice. It remains only to show that these are indeed the expected characters via our conjecture. First, note that the dimensions of the $\chi^{r}_{\bar{Q}}$ characters are given by eq. (\ref{2.-1}),
\begin{eqnarray}\label{4.10}
\frac{1}{4}\bar{Q}^2=\frac{1}{4}\sum_{i,j=1}^{r-1}\bar{Q}_iA_{ij}\bar{Q}_j=\frac{1}{2}\sum_{i=1}^{r-1}(\bar{Q}_i^2-\bar{Q}_i\bar{Q}_{i+1}),
\end{eqnarray}
with $\bar{Q}_r=0$.
Next, note that up to an integer we can write this sum using $Q$ and the $SO(2r)$ Cartan matrix,
\begin{eqnarray}\label{4.11}
\frac{1}{2}\sum_{i=1}^{r-1}(\bar{Q}_i^2-\bar{Q}_i\bar{Q}_{i+1})=\frac{1}{4}\sum_{i,j=1}^{r}Q_iD_{ij}Q_j \mod 1.
\end{eqnarray}
Which is indeed the conjectured fractional dimension, $h^{0}_Q$ (eq. \ref{-2.3}), for the $\Lambda=0$ diagram. This result along with the results for the zero momenta characters of the previous section establish our conjecture for $H^{0}_{Q}$. \\
To complete the proof of the conjecture we are left with the $\Lambda=\omega_1$ diagram.
Specifically, for the $H^{\omega_1}_{Q}$ character we conjecture the diagrammatic identity,
\begin{eqnarray}\label{4.12}
\includegraphics[width=0.38\linewidth,keepaspectratio=true]{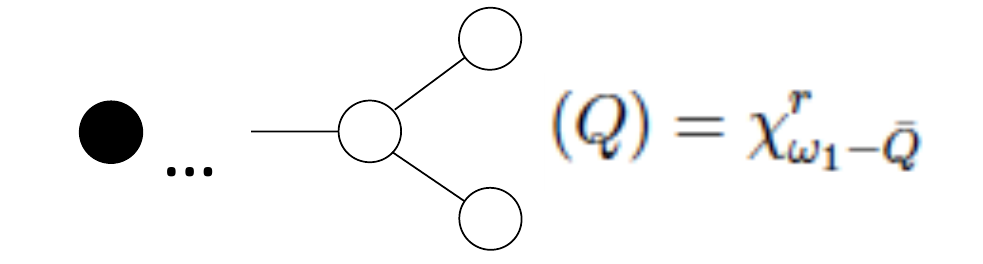}
\end{eqnarray}
where here we consider the $SO(2r)$ root $Q=(Q_1,...,Q_{r-1},Q_{r})$ while $\bar{Q}$ is a root of $SU(r)$ defined as in eq. (\ref{4.8}) and $\omega_1$ is the simple weight of $SU(r)$.
Actually, this diagram was already calculated as it is given by eq. (\ref{3.36}) with $a=0$ and $b=1$,
\begin{eqnarray}\label{4.15}
D_r(\Lambda)(Q_++Q_-)=\sum^{\infty}_{\{n_i\}=-\infty    \atop n_1 \in Z, n_i \in 2Z+\bar{Q}_i}
\frac{q^{\frac{1}{4}n^2-\frac{1}{2}n_1}}{2(q)^{r-1}_{\infty}}\sum_{s_1=\pm1}s_1^{Q_1+n_1}(1+s_1q^{n_2/2}).
\end{eqnarray}
Summing over $s_1$ dependent terms we find,
 \begin{eqnarray}\label{4.17}
\sum_{s_1=\pm1}s_1^{Q_1+n_1}(1+s_1q^{n_2/2})=
(1+(-1)^{Q_1+n_1})+q^{n_2/2}(1-(-1)^{Q_1+n_1}).
\end{eqnarray}
Note, that upon summing over $n_1$, the first term here sets $n_1 \in Q_1+2Z$ while the second sets $n_1 \in Q_1-1+2Z$. Thus, to write the summation over $n_1$ as a sum under the restriction $n_1 \in Q_1 +2Z$, simply shift $n_1 \rightarrow n_1+1$ at the second term,
 \begin{eqnarray}\label{4.18}
\sum^{\infty}_{n_1=-\infty   \atop n_1 \in 2Z+Q_1-1}   q^{(n_1^2-n_1n_2-n_1+n_2)/2}
=
\sum^{\infty}_{n_1=-\infty   \atop n_1 \in 2Z+Q_1}   q^{(n_1^2-n_1n_2+n_1)/2}.
\end{eqnarray}
Using this result, we arrive at the following identity,
 \begin{eqnarray}\label{4.20}
D_r(\Lambda)(Q_++Q_-)=\sum^{\infty}_{\{n_i\}=-\infty    \atop n_i \in 2Z+\bar{Q}_i}
\frac{q^{\frac{1}{4}n^2}}{(q)^{r-1}_{\infty}}(q^{-\frac{1}{2}n_1}+q^{\frac{1}{2}n_1})
\end{eqnarray}
where the summation is taken over $n_i=\bar{Q}_i \mod2$ for $i=1,...,r-1$. To conclude, by taking $n \rightarrow -n$, note that the two sums appearing here are equal thus we find, 
 \begin{eqnarray}\label{4.21}
\includegraphics[width=0.52\linewidth,keepaspectratio=true]{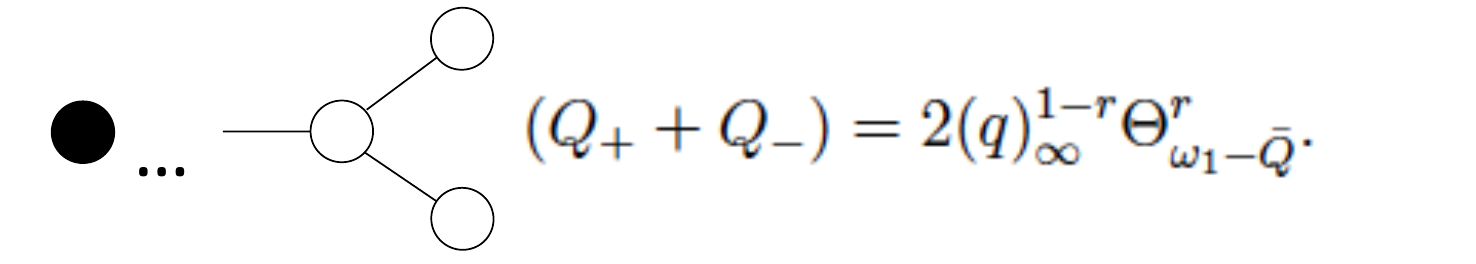}
\end{eqnarray}
Motivated by the $\Lambda=0$ diagram, let us show that the two diagrams appearing here are equivalent,
\begin{equation}\label{4.22}
 \includegraphics[width=0.8\linewidth,keepaspectratio=true]{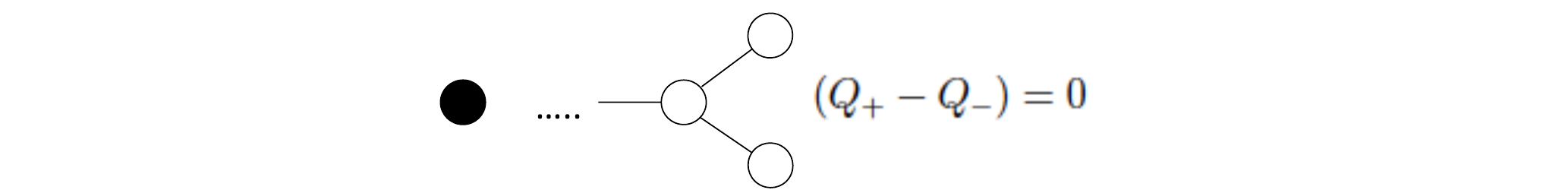}
 \end{equation}
Consider $Q_{r-1}=1$, once again, $Q_{\pm}$ transform into each other via the tail symmetry while $\omega_1$ is invariant. Thus, for $Q_{r-1}=1$, these identities are a trivial consequence of the $\Lambda=\omega_1$ diagram symmetry.\\
To prove the identities corresponding to $Q_{r-1}=0$ construct the $Q$ equivalent representation $\lambda^r_+$. For our current diagram $\lambda_+$ is given by,
\begin{eqnarray}\label{4.23}
\lambda_{+}=\omega_1+Q_{+} \mod 2P.
\end{eqnarray}
Clearly, the only difference relative to eq. (\ref{4.4}) lies in $\lambda_1$.
To facilitate the proof recall our analogous discussion above. Assuming $Q_i \neq 0$ for some $i=2,...,r-2$, this discussion was independent of $\lambda_1$ thus trivially goes through to the case at hand. We are thus left to consider the case,
\begin{eqnarray}\label{4.24}
Q_{+}=(Q_1,0,...,0),   \Rightarrow     \lambda_{+}=(1,Q_1,...,0) \mod 2.
\end{eqnarray}
 Which are again of the form studied in our discussion above. To conclude, we find the following weight belonging to the $\lambda^+_r$ equivalent $Q$ representation,
 \begin{eqnarray}\label{4.25}
\lambda_{+}-\alpha_{r-1}-\alpha_{r} \mod 2P.
\end{eqnarray}
For which the corresponding root is just $Q_{-}$, thus proving eq. (\ref{4.22}).  \\
Having established eq. (\ref{4.21}) as well as eq. (\ref{4.22}), we have proven the conjectured identity, eq. (\ref{4.12}),
corresponding to the $\Lambda=\omega_1$ diagram. For completeness let us show that the fractional dimensions of the $\Lambda=\omega_1$ diagrams indeed matches the expected dimensions following our conjecture. The dimensions corresponding to $\chi^{r}_{\omega_1-\bar{Q}}$ are given by,
 \begin{eqnarray}\label{4.26}
\frac{1}{4}(\bar{Q}-\omega_1)^2=\frac{1}{4}\sum^{r-1}_{i,j=1}\bar{Q}_iA_{ij}\bar{Q}_j-\frac{1}{2}Q_1+\frac{r-1}{4r}.
\end{eqnarray}
On the other hand, following our conjecture we have,
 \begin{eqnarray}\label{4.27}
h^{\omega_1}_{Q}=\frac{\omega_1(\omega_1+2\rho)}{4r}-\frac{1}{4}(\omega_1-Q)^2=\frac{r-1}{4r}+\frac{1}{2}Q_1-\frac{1}{4}\sum^{r}_{i,j=1}Q_iD_{ij}Q_j
\end{eqnarray}
Where $\rho=\sum_{i}\omega_i$ is the Weyl vector and $\omega_i$ are the simple weights of $SO(2r)$.
Finally, recall that $\frac{1}{4}Q^2=\frac{1}{4}\bar{Q}^2$ mod $1$, thus we find the expected fractional dimensions for the $\Lambda=\omega_1$ diagrams. \\
Having considered the $Q$ equivalent representations of $\lambda_+=b\omega_1-Q_+$ for both $b=0,1$ we find the following identity for any $Q$ and positive $b$,
\begin{equation}\label{4.28}
 \includegraphics[width=0.7\linewidth,keepaspectratio=true]{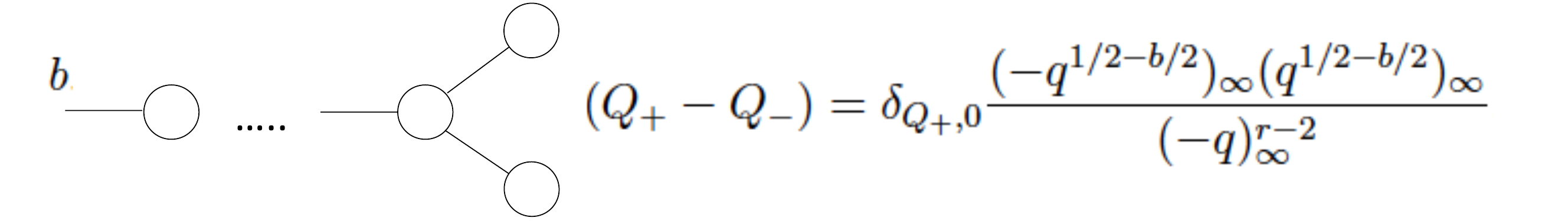}
 \end{equation}
where $Q_{\pm}$ are defined as follows. First, define $Q_1=Q$ and $Q_2=Q+\alpha_{r-1}+\alpha_r$ then,
\begin{equation}\label{4.29}
 Q_{\pm}=Q_i  \hspace{0.2in} \text{if}  \hspace{0.2in}     \frac{1}{2}Q_i^2+b\omega_1Q_i=\frac{1}{2}Q_{\pm}^2+b\omega_1Q_{\pm}  \mod2.
 \end{equation}
This concludes our proof for all the characters corresponding to fundamental weights of mark one. Finally, as in previous cases, the diagrammatic recursion relations appearing here are a broader mathematical statement. Indeed, from the simplest case of the diagrammatic identities found above, i.e $n_r=0$ and $b=0,1$, we have proven all the   characters identities corresponding to our conjecture. As we will show in the next section, these diagrammatic identities actually produce the characters 
 for $\Lambda$ any $SO(2r)$ DHW of level $2$ and $\lambda$ any weight of $SO(2r)$.
Thus, generalising our results to any character of the $SO(2r)/U(1)^r$ theory.

\section
{\large{Generalizing the conjecture}}

The identities proven in the previous sections give diagrammatic expressions for characters of fundamental weights of mark one, i.e $\Lambda=\omega_i$ for $i=0,1,r-1,r$. The purpose of this section is to give similar diagrammatic identities for the characters  of any level $2$ DHW.  In our recent paper \cite{gengep}, we mentioned that some of these DHWs characters can be found by the affine Lie algebras diagrams automorphisms.
More specifically, recall  the affine $SO(2r)$ Dynkin diagram,
\begin{equation}\label{5.0}
\includegraphics[width=0.38\linewidth]{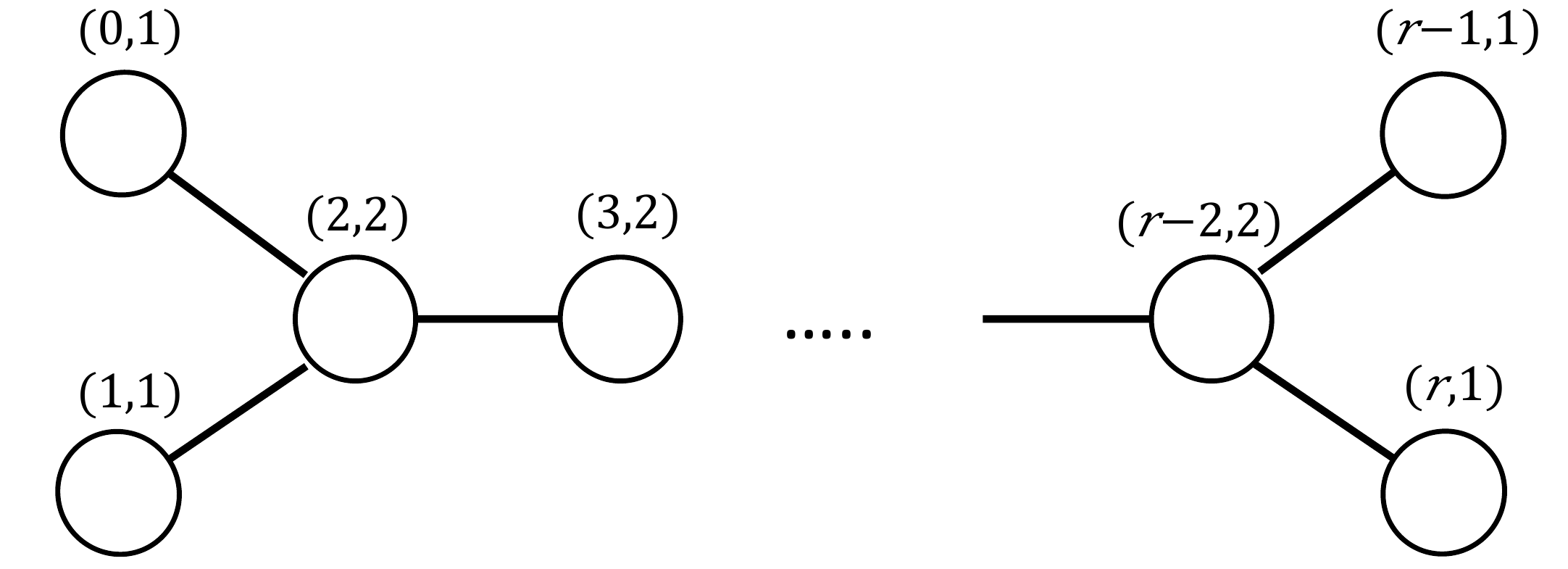}
 \end{equation}
where the parenthesis beside the nodes gives the numbering and mark of the corresponding simple root. 
The symmetry group of this diagram, denoted $O$, and its description in terms of its generating elements are given by,
\begin{equation}\label{5.1}
\nonumber
\begin{array}{lll} \hline\hline
   G & O(G) & O(G) \text{ generator action} \\ \hline 
   SO(4r) & \mathbb{Z}_2 \times \mathbb{Z}_2 & a[\Lambda_0,\Lambda_1,\Lambda_2,...\Lambda_{r-1},\Lambda_{r}]=[\Lambda_1,\Lambda_0,\Lambda_2,...\Lambda_r,\Lambda_{r-1}]  \\
     &  & \bar{a}[\Lambda_0,\Lambda_1,\Lambda_2,...\Lambda_{r-1},\Lambda_{r}]=[\Lambda_{r},\Lambda_{r-1},\Lambda_{r-2},...\Lambda_1,\Lambda_0]  \\
   SO(2+4r)  & \mathbb{Z}_4 & a[\Lambda_0,\Lambda_1,\Lambda_2,...\Lambda_{r-1},\Lambda_{r}]=[\Lambda_{r-1},\Lambda_{r},\Lambda_{r-2},...\Lambda_1,\Lambda_0] \\ \hline\hline
\end{array}
 \end{equation}
Finally, the coset characters specified by their affine extension $\hat{\Lambda}=\omega_0+\omega_i$ are symmetric under the action of $O$, 
 \begin{eqnarray}\label{5.2}
H^{\hat{\Lambda}}_{\hat{\lambda}}=H^{O\hat{\Lambda}}_{O\hat{\lambda}},
\end{eqnarray}
and we can identify all the characters belonging to some DHW orbit.
This symmetry can be used to find some of the missing DHWs characters. At level two, the DHWs are either some level $2$ combination of the four external nodes, $\hat{\Lambda}=\omega_i+\omega_j$ such that $i,j=0,1,r-1,r$ or one of the inner nodes, $\hat{\Lambda}=\omega_i$ such that $i=2,...,r-2$. Actually,  any DHW corresponding to the external nodes lies in one of the orbits corresponding to $\omega_0+\omega_i$.
It follows that the characters for all DHWs corresponding to  external nodes are identified as follows,
 \begin{eqnarray}\label{5.3}
H^{\omega_j+\omega_k}_{\lambda}=H^{\omega_0+\omega_i}_{O\lambda},
\end{eqnarray}
Where $j,k,i=0,1,r-1,r$ and $O$ is defined via,
 \begin{eqnarray}\label{5.1}
O(\omega_j+\omega_k)=\omega_0+\omega_i.
\end{eqnarray}
This identification simply means that characters corresponding to $\Lambda=\omega_i+\omega_j$ are given by one of the diagrams discussed in the previous sections with an appropriate shift of $Q$. \\
To complete the identification it remains to give a diagrammatic expression for characters corresponding to the inner nodes, i.e $\Lambda=\omega_i$ where $i=2,3,...,r-2$. To identify the corresponding characters of the $\mathbb{Z}_2$ orbifolded bosons theory let us first calculate the fractional dimensions corresponding to $\Lambda=\omega_i$, where $i=1,...,r-2$, and an arbitrary root $\beta$,
\begin{eqnarray}\label{5.4}
h^{\omega_i}_{\beta}=\frac{\omega_i(\omega_i+\rho)}{4r}-\frac{1}{4}(\omega_i+\beta)^2=\frac{i(r-i)}{4r}-\frac{1}{2}\beta_i-\frac{1}{4}\beta^2
\end{eqnarray}
To find the corresponding bosonic character consider $\bar{\lambda}=\omega_i+\bar{\beta}$, where here $\omega_i$ are the $SU(r)$ simple weights and $\bar{\beta}=(\beta_1,...,\beta_{r-2},\bar{\beta}_{r-1})$ is a root of $SU(r)$, such that $\bar{\beta}_{r-1}=\beta_r+\beta_{r-1} \mod 2$. The dimension corresponding to $\chi^r_{\omega_i-\bar{\beta}}$ is given by, 
\begin{eqnarray}\label{5.5}
\frac{1}{4}(\omega_i-\bar{\beta})^2=\frac{i(r-i)}{4r}-\frac{1}{2}\bar{\beta}_i+\frac{1}{4}\bar{\beta}^2
\end{eqnarray}
Which has exactly the same fractional part as $h^{\omega_i}_{\beta}$. Following our discussion in section ($4$) these characters are given by,
\begin{equation}\label{5.6}
\chi^r_{\omega_i-\bar{\beta}}=  (q)_{\infty}^{-r+1}\Theta^r_{\omega_i-\bar{\beta}},
\end{equation}
for $i \neq 0$. 
To identify the diagrammatic expressions corresponding to these characters let us write the characters in terms of $\psi_{a,\bar{Q}}$. First, to find $a$ and $\bar{Q}$, decompose $\bar{\lambda}$ according to eq. (\ref{3.19}),
\begin{equation}\label{5.7}
\omega_i-\bar{\beta}=\bar{Q}-a\omega_{r-1}.
\end{equation}
Multiplying this equation by $\omega_l$ we find,
\begin{equation}\label{5.8}
\bar{Q}_l=Min(l,i)-\frac{l}{r}(i+a)-\bar{\beta}_l.
\end{equation}
This determines both $\bar{Q}_l$ and $a$,
\begin{equation}\label{5.9}
a=-i+rk,   \hspace{0.2in}
\bar{Q}_l=Min(l,i)-\bar{\beta}_l-lk,
\end{equation}
up to an integer $k$. 
 It follows, from the definition of $\psi_{a,\bar{Q}}$ (eq. \ref{3.18a}), that the character corresponding to $\bar{\lambda}=\omega_i-\bar{\beta}$ is given by,
 \begin{eqnarray}\label{5.10}
\chi^r_{\omega_i-\bar{\beta}}=q^{\Delta_{i}}\psi^r_{a,\bar{Q}}
\end{eqnarray}
Where $\bar{Q}$ and $a$ are given above. As our main issue of interest are the $SO(2r)/U(1)^r$ characters, let us give an equivalent definition in the language of the $SO(2r)$ algebra.  First, recall the definition of $\bar{Q}$,
 \begin{eqnarray}\label{5.11}
\bar{Q}=(Q_1,...,Q_{r-2},Q_{r-1}+Q_r)
\end{eqnarray}
Clearly, $\psi_{a,\bar{Q}}$ is independent of $Q_{r-1}-Q_r$ (see eq. \ref{3.28}). Subsequently, let us define $Q$ as,  
\begin{equation}\label{5.12}
Q=\lambda-k(\omega_{r-1}+\omega_r)+\frac{1}{2}(\alpha_{r-1}-\alpha_r)(Q_{r-1}-Q_r+\beta_{r-1}-\beta_r),
\end{equation}
where we define $\lambda=\omega_i-\beta$.
It is easily verified that the corresponding $\bar{Q}$ is given by eq. (\ref{5.9}), moreover, the freedom to determine $Q_{r-1}-Q_r$ is manifest.
The reader might note that this definition allows for non integer $Q_{r-1}$ and $Q_r$. However, this is irrelevant as $\bar{Q}_{r-1}=Q_{r-1}+Q_r$ is clearly an integer. Finally, following eq. (\ref{3.24}), the characters $H^{\omega_i}_{\beta}$ for $i=1,...r-2$ are given by,
\begin{eqnarray}\label{5.13}
\includegraphics[width=0.55\linewidth,keepaspectratio=true]{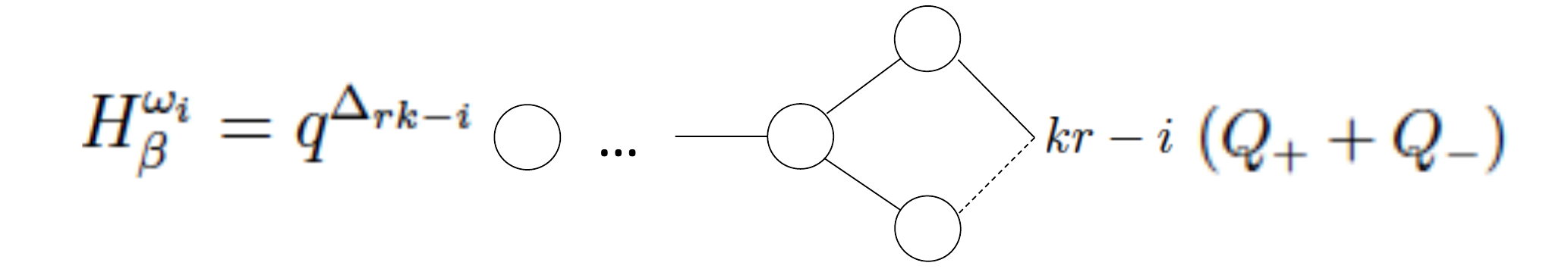}
\end{eqnarray}
with $Q_+=Q$ and $Q_{-}=Q+\alpha_{r-1}-\alpha_{r}$ and $\Delta_a=a^2(r-1)/4r+(1-r)/24$. \\
Let us take a moment to reflect on the symmetries of $H^{\omega_i}_{\beta}$. Consider the free integer $k$, different values represent different representation of the same bosonic character $\chi^r_{\omega_i-\bar{\beta}}$ thus $H^{\omega_i}_{\beta}$ is independent of $k$.
This symmetry changes $i \rightarrow i-r$ as well as $Q  \rightarrow Q-\omega_{r-1}-\omega_{r}$ and is tantamount to the outer automorphism reversing the inner nodes of the affine $SO(2r)$ diagram. This is expected as this symmetry is inherited from the $SU(r)$ theta function. Next, let us consider the $Q$ symmetry eq. (\ref{1.6.1}) of section ($3$). This symmetry implies $H^{\omega_i}_{\beta}$ is symmetric under the Weyl generators $s_i$, for $i=1,...,r-2$, as clearly the $Q$ symmetry of the sub-diagram containing only the first $r-2$ nodes is still viable. More specifically, first to ease the discussion let us set $k=0$. Then for the diagram in question we find the symmetry,
 \begin{eqnarray}\label{5.14}
Q_i \rightarrow Q_i+\lambda_i-(\delta_{i,r-1}-\delta_{i,r})(i+Q_{r-1}-Q_r+\beta_{r-1}-\beta_r),
\end{eqnarray}
by noting that the diagram external momenta is specified by $-i(\omega_{r-1}-\omega_{r})$ and using the above definition of $Q$. Clearly, for $i=1,...,r-2$, this reproduces the transformation under the Weyl generators $s_i$ of eq. (\ref{1.2}). Finally, the symmetry under the last two generators doesn't follow from our discussion on $Q$ symmetry, rather, it is a simple consequence of the summation over $Q=Q_{\pm}$. Thus, our expression for $H^{\omega_i}_{\beta}$ is also symmetric under the action of the Weyl Group. Finally, note that changing $i \rightarrow -i$ interchanges $Q_{r-1} \leftrightarrow Q_r$ which is also a symmetry due to $Q$ the summation. \\
To proceed as in previous cases, one would like to argue that the diagrams appearing in eq. (\ref{5.13}) are equivalent. However, this argument relied on  $Q$ symmetry valid for diagrams including only non negative external momenta. We can retrieve this symmetry in the following manner. First, denote the diagram appearing in eq. (\ref{5.13}) by $D(\Lambda,Q)$. Using the $i \leftrightarrow -i$ symmetry,
 \begin{eqnarray}\label{5.15}
H^{\omega_i}_{\beta}=\frac{1}{2}D(\Lambda,Q_+)+\frac{1}{2}D(-\Lambda,Q_+) +\frac{1}{2}D(\Lambda,Q_-)+\frac{1}{2}D(-\Lambda,Q_-).
\end{eqnarray}
Next, define the $\Lambda \rightarrow -\Lambda$ symmetric diagram $\tilde{D}(\Lambda,Q)=D(\Lambda,Q)+D(-\Lambda,Q)$. Now consider $\tilde{D}(\Lambda,Q_+)$, changing $\Lambda \rightarrow -\Lambda$ interchanges $Q_{r-1} \leftrightarrow Q_r$. By setting $Q_{r-1}-Q_r=1 \mod2$ the $\Lambda$ symmetry interchanges $Q_+ \leftrightarrow Q_-$. For example, consider the case $Q_{r-1}+Q_{r}=1$, setting $Q_{r-1}-Q_r=1$ we find $Q_+=(Q_1,....,Q_{r-2},1,0)$, indeed, changing $Q_{r-1} \leftrightarrow Q_r$ we find $Q_{-}$. Alternatively, consider $Q_{r-1}+Q_{r}=0$ for which we find $Q_+=(Q_1,....,Q_{r-2},1/2,-1/2)$. Again, changing $Q_{r-1} \leftrightarrow Q_r$ we find $Q_{-}$ to conclude 
 \begin{eqnarray}\label{5.16}
\tilde{D}(\Lambda,Q_+)=\tilde{D}(\Lambda,Q_-).
\end{eqnarray}
So that all the characters corresponding to $\omega_i$ for $i=1,...,r-2$ are given by,
\begin{eqnarray}\label{5.17}
\includegraphics[width=0.57\linewidth,keepaspectratio=true]{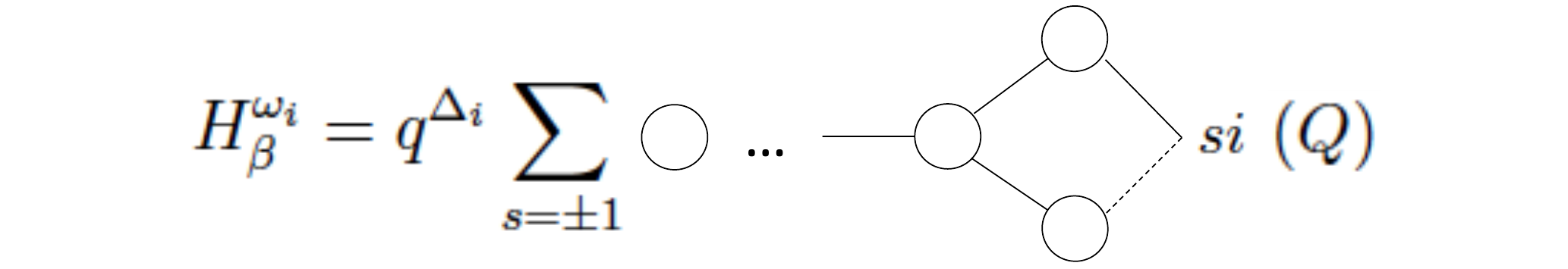},
\end{eqnarray}
where $\lambda=\omega_i-\beta$ and $Q$ is given by
 \begin{eqnarray}\label{5.18}
Q=\lambda+\frac{1}{2}(\alpha_{r-1}-\alpha_r)(1+\beta_{r-1}-\beta_r). 
\end{eqnarray}

\section
{\large{Summary}}

Let us take a moment to summarise our results, consider
\begin{equation}\label{-1.4}
 \includegraphics[width=0.5\linewidth,keepaspectratio=true]{1}
 \end{equation}
Then for the following cases these diagrams were shown to be given by,
\begin{enumerate}
  \item $\Lambda=b\omega_1 $:
     \begin{eqnarray}
     \nonumber
        D_r(\Lambda)(Q_+-Q_-)= \frac{\delta_{Q_+,0}}{(-q)^{r-2}_{\infty} }     (q^{1-b},q^2)_{\infty}.
     \end{eqnarray}  
   \item  $\Lambda=b\omega_1+a(\omega_{r-1}-\omega_{r})$:
     \begin{eqnarray}
     \nonumber
       D_r(\Lambda)(Q_++Q_-)=\sum^{\infty}_{\{n_i\}=-\infty    \atop {n_1 \in Z \atop{ n_i \in 2Z+Q_i \atop n_{r-1} \in 2Z+Q_{r-1}+Q_r}}}
       \frac{q^{\frac{1}{4}n^2-\frac{1}{2}n_1b-\frac{1}{2}n_{r-1}a}}{2(q)^{r-1}_{\infty}}\sum_{s_1=\pm1}s_1^{Q_1+n_1} \frac{
       (-s_1q^{(1-b+n_2)/2})_{\infty}}{(-s_1q^{(1+b+n_2)/2})_{\infty}}.
     \end{eqnarray}
   \item $\Lambda=b\omega_1+\omega_r$:
     \begin{eqnarray}
     \nonumber  
         D_r(\Lambda)(Q_++s_rQ_-)=\frac{1}{2}  \sum_{s_{1}}   \frac{  (-s_{1}q^{1/2-b/2})_{\infty} (-s_{1}s_rq^{-b/2})_{\infty} } {(s_rq^{1/2})^{r-2}_{\infty}}.
     \end{eqnarray}
\end{enumerate} 
Where for any $Q$ we define $Q_1=Q$ and $Q_2=Q+\alpha_{r-1}+\alpha_r$ then $Q_{\pm}$ are determined as follows,
\begin{equation}\label{4.29}
 Q_{\pm}=Q_i  \hspace{0.2in} \text{if}  \hspace{0.2in}     \frac{1}{2}Q_i^2+\Lambda Q_i=\frac{1}{2}Q_{\pm}^2+\Lambda Q_{\pm}  \mod2.
 \end{equation}
Finally, We have shown that these diagrammatic identities, at some specific values of $a$ and $b$, produce all the characters of the level two $H(SO(2r))$ coset.
One may recall, that in section ($4$) we have argued that these series include the Ramanujan identities (\ref{2.3.10.1}) as well as some cases of the Cauchy identity (\ref{2.3.4}) and Heine's sum (\ref{3.33aa}).
However, when proving these series we have not used these identities, thus implying that one can prove them from our results. More specifically, consider the change of variables $a_i=s_iq^{-\Lambda_i/2}$ for $i=1,...r-1$ while $a_r=s_r$. When proving the second identity we did not assume that $a$ or $b$ are integers so that $a_1$ and $a_{r-1}$ are not restricted. However, for the first and third identity one may recall the use of 
 $Q$ symmetry, nonetheless these can be easily generalised in a similar manner. Clearly, if $b$ is not an integer, the $Q$ symmetry of the sub diagram excluding the first node is still applicable. One can simply retrieve the $Q_1$ dependance to find,
\begin{enumerate}
  \item $\Lambda=b\omega_1 $:
     \begin{eqnarray}
     \nonumber
      \sum_{ \{b_i\}=0\atop }^\infty 
  \frac{ q^{ \frac{1}{4}b^2}a_1^{b_1}a_2^{b_2}...a_{r-1}^{b_{r-1}-b_r}(-1)^{b_r}} {(q)_{b_1}(q)_{b_2}...(q)_{b_r}}  =
    \frac{1}{(-q)^{r-2}_{\infty} }     (qa^2_1,q^2)_{\infty}.
     \end{eqnarray}  
   \item  $\Lambda=b\omega_1+a(\omega_{r-1}-\omega_{r})$:
     \begin{eqnarray}
     \ \nonumber
     \sum_{ \{b_i\}=0\atop }^\infty 
  \frac{ q^{ \frac{1}{4}b^2}a_1^{b_1}a_2^{b_2}...a_{r-1}^{b_{r-1}-b_r}} {(q)_{b_1}(q)_{b_2}...(q)_{b_r}}  =
    \sum^{\infty}_{\{n_i\}=-\infty} \frac{q^{\frac{1}{4}n^2}a_1^{n_1}a_2^{n_2}...a_{r-1}^{n_{r-1}}}{(q)^{r-1}_{\infty}} \frac{
       (-a_1q^{(1+n_2)/2})_{\infty}}{(-q^{(1+n_2)/2}/a_1)_{\infty}}
     \end{eqnarray}
   \item $\Lambda=b\omega_1+\omega_r$:
     \begin{eqnarray}
    \nonumber
 \sum_{a_2,...a_{r-1}}     \sum_{ \{b_i\}=0\atop }^\infty 
  \frac{ q^{ \frac{1}{4}b^2-\frac{1}{2}b_r}a_1^{b_1}a_2^{b_2+Q_2}...a_{r-1}^{b_{r-1}-b_r+Q_{r-1}-Q_r}a_r^{b_r}} {2^{r-2}(q)_{b_1}(q)_{b_2}...(q)_{b_r}} =a_r^{\frac{1}{2}Q^2} \frac{  (-a_{1}a_r^{Q_2}q^{1/2})_{\infty} (-a_{1}a_r^{1+Q_2})_{\infty} } {(a_rq^{1/2})^{r-2}_{\infty}}.
     \end{eqnarray}
\end{enumerate} 
Here, we first realise the parity restriction using $s_{r}^{b_r+Q_r}\sum_{s_1,...s_{r-1}=\pm1}s_1^{b_1+Q_1}...s_{r-1}^{b_{r-1}-b_r+Q_{r-1}-Q_r}$ and replace $a_i=s_iq^{-\Lambda_i/2}$ for $i=1,...r-1$ while $a_r=s_r$. Next, we sum over $\sum_{\bar{Q}_i}a_i^{\bar{Q}_i}$ with $i=1...r-1$ for the first two identities and $i=1$ for the third, where recall $\bar{Q}_{r-1}=Q_{r-1}+Q_r$. When using these identities one should be careful as the values of the $a_i$'s are restricted. For example, in the third identity $a_1$ can be chosen freely, however, $a_i=\pm1$ for $i=2,...,r$. \\
Next consider the second identity, this was proven using the Jacobi identity to solve $D_2$ which was our initial condition while Heine's sum for $D_3$ provided the appropriate recursion relation. To find the diagrammatic recursion relation $D_3$ was solved by first summing over the first node. To prove the first Ramanujan identity, corresponding to $D_3$ at $b=0$, consider first summing over the second and third nodes, replacing $b_1=2n_1+Q_1$ and extracting the $n_1$ dependance,
\begin{eqnarray}
     \nonumber
     \sum_{Q_2,Q_3=0,1} a_2^{Q_2+Q_3}D_{3}(\Lambda,Q)= A\sum_{n_1=0 }^\infty \frac{q^{n_1(n_1+Q_1)}}{(q)_{2n_1+Q_1}}
       (-a_2q^{(1-Q_1)/2})_{n_1+Q_1}(-q^{(1+Q_1)/2}/a_2)_{n_1},
\end{eqnarray}
\begin{eqnarray}
A=a_2^{Q_1}q^{Q_1^2/2}(-a_2q^{(1-Q_1)/2})_{\infty}(-q^{(1+Q_1)/2}/a_2)_{\infty}.
\end{eqnarray}
Where in addition we sum over $a_2^{Q_2+Q_3}$, where $a_2=\pm1$, to simplify the result. It is immediately apparent that for $Q_1=b=0$ the summation appearing here coincides with the RHS of the first Ramanujan identity of eq. (\ref{2.3.10.1}). On the other hand, this is given by the second identity above, specifically for $b=0$
\begin{eqnarray}
\sum^{\infty}_{\{n_i\}=-\infty    \atop {n_1,n_2 \in Z }}
    \sum_{a_1=\pm1}   \frac{q^{\frac{1}{2}n_1^2-\frac{1}{2}n_1n_2+\frac{1}{2}n_2^2}}{2(q)^{2}_{\infty}}a_2^{n_2}a_1^{Q_1+n_1}=\frac{Aa_2^{-Q_1}}{(q)_{\infty}}\sum_{n_1=-\infty }^{\infty}q^{\frac{3}{2}n_1(n_1+Q_1)}a_2^{n_1}
\end{eqnarray}
where to solve the sum over $n_2$ we use the Jacobi identity (\ref{2.3.10}) and then extract the $n_1$ dependance in the usual way. Indeed, for $Q_1=0$ we prove the first Ramanujan identity, perhaps more interesting is that this identity also holds for any integer $Q_1$,
\begin{eqnarray}
   \frac{(q^3,aq^{3(1+Q)/2},a^{-1}q^{3(1-Q)/2};q^3)_{\infty}}{(q)_{\infty}}  =(-a)^Q\sum_{n=0 }^\infty \frac{q^{n(n+Q)}}{(q)_{2n+Q}}
       (aq^{(1-Q)/2})_{n+Q}(a^{-1}q^{(1+Q)/2})_{n}.         
\end{eqnarray}
where we replace $a_2 \rightarrow -a$ and use the Jacobi identity. In a similar manner, one can prove the second Ramanujan identity of eq. (\ref{2.3.10.1}) as well as some extension for other values of $Q_2$ by considering the $r=4$ diagram of the second identity above and preforming the sum over the external nodes.  \\
Regarding the first and third identities, when proving these we have used only the Euler identity (\ref{1.4}). Recall, that in section ($4$) we have shown $D_3$, of the first identity above, corresponds to the Cauchy identity (\ref{2.3.4}) at $z=0$. Having proven the identity for $D_3$ with no use of the Cauchy identity one can regard the correspondence of section ($4$) as a proof for the $z=0$ Cauchy identity. In a similar fashion, one can prove the Cauchy identity at $z=\pm q^{1/2}$ using our results for $D_3$ of the third identity above. Moreover, one can also consider the identities arising from $D_4$, of the first or third type, for example consider, 
\begin{eqnarray}
     \sum_{Q_1=0,1 \atop
     Q_3+Q_4=0,2}  a_1^{Q_1}a_4^{(Q_3+Q_4)/2}D_{4}(\Lambda,Q)= \sum_{a_2,a_3=\pm1}\sum_{\{n_i\}=0 }^\infty \frac{q^{n^2/4-n_4/2}a_1^{n_1}a_2^{n_2+Q_2}a_3^{n_3+n_4}a_4^{n_4}}{4(q)_{n_1}(q)_{n_2}(q)_{n_3}(q)_{n_4}},
\end{eqnarray}
where $\Lambda=b\omega_1+\omega_4$, $Q=(Q_1,Q_2,Q_3,Q_4)$ and $a_i$'s are defined above. Let us first consider the last two nodes, 
\begin{eqnarray}
     N_3N_4=\sum_{a_3}(-a_3q^{(1-n_2)/2})_{\infty}(-a_3a_4q^{-n_2/2})_{\infty}=(-y^{-n_2},y)_{\infty}
\end{eqnarray}
where $y=a_4q^{1/2}$ and we note that the contribution from $a_3a_4^{1-n_2}=-1$ vanishes. Solving for the first node and extracting the $n_2$ dependance,
\begin{eqnarray}
     N_1=(-a_1a_4^{1-Q_2}y^{1-Q_2},y^2)_{\infty}/(-a_1a_4^{1-Q_2}y^{1-Q_2},y^2)_{-(n_2-Q_2)/2}.
\end{eqnarray}
To simplify the result we define $y^2/a=a_1a_4^{1-Q_2}y^{1-Q_2}$. In terms of $y$ and $a$ one finds,
\begin{eqnarray}
     \nonumber
     \sum_{Q_1=0,1 \atop
     Q_3+Q_4=0,2}  s_1^{Q_1}s_4^{(Q_3+Q_4)/2}D_{4}(\Lambda,Q)= A\sum_{n_2=0 \atop n_2=Q_2 \mod2}^\infty \frac{y^{n_2(n_2+2Q_2)/4}(-a,y^2)_{(n_2-Q_2)/2} }{(y)_{n_2}a^{(n_2-Q_2)/2}}
\end{eqnarray}
\begin{eqnarray}
A=a_4^{Q_2}y^{-Q_2(Q_2+2)/4}(-y,y)_{\infty}(-y^2/a,y^2)_{\infty}.
\end{eqnarray}
On the other hand, this diagram is given by the third identity above. Writing it in terms of $y$ and comparing we find,
\begin{eqnarray}
  \sum_{n=0 \atop n=Q \mod2}^\infty \frac{y^{n(n+2Q)/4}(-a,y^2)_{(n-Q)/2} }{(y)_{n}a^{(n-Q)/2}}
 =
      \frac{ y^{Q(Q+2)/4} (-y/a,y^2)_{\infty}} {(-y/a)_{Q}(y,y^2)_{\infty}}.
\end{eqnarray}
which holds for any $a$ and integer $Q$. Indeed, one can easily verify, that for $Q=0$, going back to $q$ one finds Heine's sum (\ref{3.33aa}) for general $a$ and $c=a_4q^{1/2}$.

\section
{\large{Discussion }}

Perhaps, the most intriguing part of this work, is the appearance of the extended identities and their interpretation as $q$-diagrams. 
Our main objective was to prove our conjecture for the level two $H(SO(2r))$ coset characters corresponding to  $\Lambda$ zero or a fundamental weight of mark one. As discussed, these identities indeed provide us with all the $q$-diagrams corresponding to our conjecture. It's interesting to note that using these diagrammatic identities we generalised our conjecture. More specifically, an $SO(2r)$ diagram was found for all the characters of $H(SO(2r))$ at level two. Thus, all the $SO(2r)$ level $2$ string functions were found to be given by diagrammatic expressions. These results raise two interesting questions regarding generalisations to other simply laced Lie algebras. First, as was discussed in section ($3$), the basic $Q$ symmetry of $q$-diagrams holds for all Lie algebras. Furthermore, the language of $q$ diagrams makes apparent the recursive nature of our conjecture. It is thus somewhat tempting to use $q$-diagrams to prove our conjecture for all simply laced Lie algebras. Indeed, we intend to publish a similar proof for the $SU(r)$ conjecture in the near future \cite{gengep1}. The second question, is whether all of the $H(G_r)$ coset characters correspond to some $G_r(\Lambda,Q)$, i.e a $G_r$ diagram along with some weight $\Lambda$ and root $Q$. \\
Regarding number theory and in particular $q$ identities it is interesting that many known identities can be associated with $q$-diagrams. Finding the $q$-diagrams corresponding to these identities revealed an intriguing relation between them. More specifically, they correspond to the first terms of an infinite series of diagrammatic identities which in turn means that all the identities in the series follow from the first identity and an appropriate recursion relation. Furthermore, many more identities can be proven by simple manipulations as changing the order of nodes summation or using $Q$ symmetry. \\
To conclude, there remain some open questions to address regarding these identities and $q$-diagrams in general. First, many identities are somewhat hidden in the mentioned series in a similar way to the Euler, Cauchy, Heine, Jacobi and Ramanujan identities. A complete description of those lies beyond the scope of our current work. 
Additionally, one can generalise our results by considering novel sum restrictions such as $b_i=Q_i \mod k$ for any integer $k$.
Second, during our study we have encountered many identities which can be given a diagrammatic interpretation. Although our current interest lies in Lie algebra $q$-diagrams these are but a small subspace and one can study the entire space of $q$-diagrams. Indeed, our work can be implemented to any $q$-diagram which includes as a sub diagram one of the diagrams studied here (eq. \ref{-1.4}). For example one may consider the $q$-diagrams corresponding to the affine $SO(2r)$ Dynkin diagram (eq. \ref{5.0}). Another interesting question is the study of these identities in the context of Bailey pairs \cite{bai}. Furthermore, as these characters produce the $SO(2r)$ level two string functions our results can be applied to the study of the $SO(2r)$ algebra. Finally, as mentioned in the introduction, the character identities for the $H(G_r)$ coset are closely related to some RSOS models, one can hope that using the identities for the $H(SO(2r))$ coset the relevant RSOS model can be identified and solved. From this point of view a natural question is whether the correspondence between RSOS, CFT and $q$-diagrams goes beyond simply laced Lie algebra $q$-diagrams.

\end{document}